\documentclass[pre,twocolumn,preprintnumbers,amsmath,amssymb,nofootinbib,floatfix]{revtex4}

\usepackage{graphicx,bm}

\makeatletter
\def\graphicscale{\twocolumn@sw{0.3}{0.4}}
\def\graphicthreescale{\twocolumn@sw{0.3}{0.4}}

\begin{document}

\title{Three-dimensional ferromagnetic CP$^{N-1}$ models }

\author{Andrea Pelissetto}
\affiliation{Dipartimento di Fisica dell'Universit\`a di Roma Sapienza
        and INFN Sezione di Roma I, I-00185 Roma, Italy}

\author{Ettore Vicari} 
\affiliation{Dipartimento di Fisica dell'Universit\`a di Pisa
        and INFN Largo Pontecorvo 3, I-56127 Pisa, Italy}

\date{\today}

\begin{abstract}
We investigate the critical behavior of three-dimensional
ferromagnetic CP$^{N-1}$ models, which are characterized by a global
U($N$) and a local U(1) symmetry. We perform numerical simulations of
a lattice model for $N=2$, 3, and 4.  For $N=2$ we find a critical
transition in the Heisenberg O(3) universality class, while for $N=3$
and 4 the system undergoes a first-order transition. For $N=3$ the
transition is very weak and a clear signature of its discontinuous
nature is only observed for sizes $L\gtrsim 50$.  We also determine
the critical behavior for a large class of lattice Hamiltonians in the
large-$N$ limit. The results confirm the existence of a stable large-$N$ 
CP$^{N-1}$ fixed point.  However,  this evidence contradicts the standard picture
obtained in the Landau-Ginzburg-Wilson (LGW) framework using a
gauge-invariant order parameter: the presence of a cubic term in the
effective LGW field theory for any $N\ge 3$ would usually be taken as
an indication that these models generically undergo first-order
transitions.
\end{abstract}

\maketitle


\section{Introduction}
\label{intro}

CP$^{N-1}$ models are a class of quantum field theories in which the
fundamental field is a complex $N$-component unit vector ${\bm z}({\bm
  x})$, associated with an element of the complex projective manifold
CP$^{N-1}$. They are characterized by a global U($N$) symmetry
\begin{equation}
{\bm z}({\bm x}) \to U {\bm z}({\bm x})\qquad U\in {\rm U}(N),
\label{unsym}
\end{equation}
and a local U(1) gauge symmetry 
\begin{equation}
{\bm z}({\bm x}) \to e^{i\Lambda({\bm x})} {\bm z}({\bm x}).
\label{u1gausym}
\end{equation}
The corresponding statistical field theory is defined by
\begin{eqnarray}
&&{\cal Z} = \int [d{\bm z}] \ \exp \left[{-\int d{\bm x} {\cal
        L}({\bm z})}\right] , \nonumber \\ && {\cal L} = {1\over 2 g}
  \overline{D_{\mu}{\bm z}}\cdot D_\mu {\bm z}, \qquad D_\mu =
  \partial_\mu + i A_\mu, \qquad
\label{contham}
\end{eqnarray}
where $A_\mu = i\bar{{\bm z}}\cdot \partial_\mu {\bm z}$ is a
composite gauge field, which transforms as $A_\mu({\bm x}) \to
A_\mu({\bm x}) - \partial_\mu \Lambda({\bm x})$ under the gauge
transformations (\ref{u1gausym}).  For $N=2$ the CP$^1$ field theory
is locally isomorphic to the O(3) non-linear $\sigma$ model with the
identification $s^a=\sum_{ij} \bar{z}^i \sigma_{ij}^{a} z^j$, where
$a=1,2,3$ and $\sigma^{a}$ are the Pauli matrices.

Three-dimensional (3D) CP$^{N-1}$ models emerge as effective theories
of SU($N$) quantum
antiferromagnets~\cite{RS-90,TIM-05,TIM-06,SHOIM-08,Kaul-12,KS-12,BMK-13}
and of scalar electrodynamics with a compact U(1) gauge group.  The
two-dimensional (2D) CP$^{N-1}$ model is instead an interesting
theoretical laboratory to study quantum field theories of fundamental
interactions as it shares several features with quantum chromodynamics
(QCD), the theory that describes the hadronic strong
interactions~\cite{ZJ-book,MZ-03}.

The simplest lattice formulation of CP$^{N-1}$ model is obtained by a
straightforward discretization of the continuum theory
(\ref{contham}). One considers $N$-component complex unit vectors
${\bm z}_{\bm x}$ defined on the sites of a lattice and the
nearest-neighbor Hamiltonian
\begin{equation}
H_s = - J \sum_{{\bm x}\mu} | \bar{\bm{z}}_{\bm
  x} \cdot {\bm z}_{{\bm x}+\hat\mu} |^2,
\label{hcpn}
\end{equation}
where the sum is over the lattice sites ${\bm x}$ and the lattice
directions $\mu$ ($\mu = 1,\ldots,d$), and
$\hat\mu=\hat{1},\hat{2},\ldots$ are unit vectors along the lattice
directions.

In three dimensions CP$^{N-1}$ models are expected to undergo a
finite-temperature transition between a high- and a low-temperature
phase.  Its nature may be investigated by resorting to the
Landau-Ginzburg-Wilson (LGW) field-theoretical
approach~\cite{Landau-book,WK-74,Fisher-75,Ma-book,PV-02}.  In this
framework the critical features are uniquely specified by the nature
of the order parameter associated with the critical modes, by the
symmetries of the model, and by the symmetries of the phases
coexisting at the transition, the so-called symmetry-breaking pattern.
In the presence of gauge symmetries, the traditional LGW approach
starts by considering a gauge-invariant order parameter, effectively
integrating out the gauge degrees of freedom, and by constructing a
LGW field theory that is invariant under the global symmetries of the
original model.

The order parameter of the transition in ferromagnetic CP$^{N-1}$
models is usually identified with the gauge-invariant site variable
\begin{equation}
Q_{{\bm x}}^{ab} = \bar{z}_{\bm x}^a z_{\bm x}^b - {1\over N}
\delta^{ab},
\label{qdef}
\end{equation}
which is a hermitian and traceless $N\times N$ matrix. It
transforms as 
\begin{equation}
Q_{{\bm x}} \to {U}^\dagger Q_{{\bm x}} \,{U},
\label{symmetry-U(N)}
\end{equation}
under the global U($N$) transformations (\ref{unsym}).  The
order-parameter field in the corresponding LGW theory is therefore a
traceless hermitian matrix field $\Phi^{ab}({\bm x})$, which can be
formally defined as the average of $Q_{\bm x}^{ab}$ over a large but
finite lattice domain.  The LGW field theory is obtained by
considering the most general fourth-order polynomial in $\Phi$
consistent with the U($N$) symmetry (\ref{symmetry-U(N)}):
\begin{eqnarray}
{\cal H}_{\rm LGW} &=& {\rm Tr} (\partial_\mu \Phi)^2 
+ r \,{\rm Tr} \,\Phi^2 \label{hlg}\\
&& +  w \,{\rm tr} \,\Phi^3 
+  \,u\, ({\rm Tr} \,\Phi^2)^2  + v\, {\rm Tr}\, \Phi^4 .
\nonumber
\end{eqnarray}
A continuous transition is possible if the renormalization-group (RG)
flow computed in the LGW theory has a stable fixed point.

For $N=2$, the cubic term in Eq.~(\ref{hlg}) vanishes and the two
quartic terms are equivalent.  Therefore, one recovers the
O(3)-symmetric LGW theory, consistently with the equivalence between
the CP$^1$ and the Heisenberg model. For $N\ge 3$, the cubic term is
generically expected to be present.  The presence of a cubic term in
the LGW Hamiltonian is usually taken as an indication that phase
transitions occurring in this class of systems are generically of
first order.  Indeed, a straightforward mean-field analysis shows that
the transition is of first order in four dimensions.  The nature of
the transition should not change sufficiently close to four
dimensions, as long as statistical fluctuations are small. In
particular, it is usually assumed that the four-dimensional mean-field
result also applies in three dimensions.  In this scenario, continuous
transitions may still occur, but they require a fine tuning of the
microscopic parameters leading to the effective cancellation of the
cubic term. If this occurs, the critical behavior is controlled by the
stable fixed point of the RG flow of the LGW theory (\ref{hlg}) with
$w=0$. Such a LGW model was studied in detail in
Ref.~\cite{DPV-15}. For $N=3$ it is equivalent to the O(8) vector
$\phi^4$ theory, therefore there is a stable fixed point, while no
fixed point was identified for $N\ge 4$ by using field-theoretical
methods.

The prediction of a first-order transition is apparently contradicted
by recent numerical studies~\cite{NCSOS-11,NCSOS-13,Kaul-12}.
Refs.~\cite{NCSOS-11,NCSOS-13} report evidence of a continuous
transition in a loop model supposed to belong to the same universality
class as that of the 3D CP$^2$ model.  Their results hint at a new
CP$^2$ universality class,~\cite{NCSOS-13} characterized by the
correlation-length critical exponent $\nu=0.536(12)$.  This result is
not consistent with the standard LGW picture we reported above. In
the standard scenario in which generic CP$^2$ models undergo
first-order transitions due to the cubic term in the corresponding
LGW theory, a continuous transition may only emerge from
a tuning of the parameters achieving its cancellation. Therefore,
the continuous transition associated with the effective
cancellation of the cubic term
is described by the stable fixed
point of the LGW theory without cubic term, which belongs to
the O(8) vector universality class, whose critical exponent~\cite{DPV-15}
$\nu=0.85(2)$ definitely differs from that reported in
Ref.~\cite{NCSOS-13}. Therefore, the emergence of a new CP$^2$
universality class is in apparent contradiction with the predictions
of the standard LGW picture, in which the presence of a cubic term in
the effective LGW Hamiltonian implies that first-order phase
transitions.  To make the existence of the CP$^2$ universality class
plausible even in the presence of a cubic term,
Ref.~\cite{NCSOS-13} proposed a double
expansion around $N=2$ (where the cubic term vanishes) and
$\epsilon=4-d$, arguing that a continuous transition may be possible
for values of $N$ sufficiently close to $N=2$.

An additional apparent contradiction emerges when considering
CP$^{N-1}$ models in the large-$N$ limit. In this limit one may argue
that 3D CP$^{N-1}$ models may undergo a continuous transition, because
the corresponding continuum field theory is expected to share the same
critical behavior as the abelian Higgs model for an $N$-component
complex scalar field coupled to a dynamical U(1) gauge
field~\cite{MZ-03}.  This equivalence is conjectured to extend to
finite $N$ at the critical point~\cite{MZ-03}.  The RG flow of the
abelian Higgs model presents a stable fixed point for a sufficiently
large number of components~\cite{HLM-74}. More precisely, in the
perturbative $\epsilon$ expansion ($\epsilon=4-d$) one finds a stable
fixed point for $N>N_c$, with~\cite{HLM-74,MZ-03} $N_c= 90 +
24\sqrt{15} + O(\epsilon)$.  Thus, for large values of $N$, 3D
CP$^{N-1}$ models may undergo a continuous transition, in the same
universality class as that occurring in the abelian Higgs model.

The above results may suggest that the critical modes at the
transition are not exclusively associated with the gauge-invariant
order parameter $Q$ defined in Eq.~(\ref{qdef}). Other features, for
instance the gauge degrees of freedom, may become relevant, requiring
an effective description different from that of the LGW theory
(\ref{hlg}).

We mention that the LGW description apparently fails also in other
systems in which a gauge symmetry is present, for instance in
antiferromagnetic 3D CP$^{N-1}$ and RP$^{N-1}$
models~\cite{PTV-17,PTV-18}, although the discrepancies are not
related to the presence of a cubic term in the corresponding LGW
theory.

In this paper we discuss the critical behavior of ferromagnetic 3D
CP$^{N-1}$ models, presenting additional numerical and analytical
results, which may help clarify the above apparent contradictions
between the actual behavior of lattice CP$^{N-1}$ models and the LGW
approach. We consider an alternative lattice formulation of 3D
CP$^{N-1}$ models, which contains an additional U(1) gauge variable
and is linear with respect to all lattice variables, and perform
numerical simulations for $N=2$, $N=3$, and $N=4$. Moreover, we
reconsider the large-$N$ limit, presenting analytical results for the
phase diagram. Our results for $N=2,3,4$ are consistent with the standard 
LGW picture.  Indeed, while the transition is continuous for $N=2$, for
$N=3$ and 4 the transition is of first order (for $N=4$ we confirm
the results of Ref.~\cite{KHI-11}), although very weak for $N=3$.

The paper is organized as follows.  In Sec.~\ref{lattmod} we define
the 3D CP$^{N-1}$ model we consider and the observables that are
determined in the Monte Carlo simulations.  In Sec.~\ref{sumfss} we
summarize the main features of the finite-size scaling (FSS) theory at 
continuous and
first-order transitions, which allow us to distinguish the nature of
the transitions observed in lattice 3D CP$^{N-1}$ models.  Then, in
Secs.~\ref{ncp1}, \ref{ncp3}, and \ref{ncp2}, we report a numerical
study of the 3D CP$^{N-1}$ models for $N=2$, $N=4$, and finally for
$N=3$ (the most controversial case), respectively.  For $N=2$ the
transition is continuous and belongs to the O(3) vector universality
class, while it is of first order for $N=4$. For $N=3$, the transition
is also of first order, but it is so weak that its first-order nature
only emerges when relatively large systems are considered.  In
Sec.~\ref{ncpinf} we discuss the large-$N$ limit of lattice CP$^{N-1}$
models. Finally, in Sec.~\ref{conclu} we summarize our main results
and draw our conclusions. In App.~\ref{App-largeN} we discuss the
large-$N$ behavior of a general class of CP$^{N-1}$ models.

\section{Lattice three-dimensional CP$^{N-1}$ models}
\label{lattmod}

In our study we consider a lattice formulation, in which gauge
invariance is guaranteed by introducing a U(1) link variable
$\lambda_{{\bm x},\mu} \equiv e^{i\theta_{{\bm
      x},\mu}}$~\cite{RS-81,DHMNP-81,BL-81}.  The Hamiltonian is given
by
\begin{equation}
H_{\lambda} = - t N
\sum_{{\bm x}, \mu}
\left( \bar{\bm{z}}_{\bm x} 
\cdot \lambda_{{\bm x},\mu}\, {\bm z}_{{\bm x}+\hat\mu} 
+ {\rm c.c.}\right),
\label{hcpnla}
\end{equation}
where the sum is over all lattice points ${\bm x}$ and lattice
directions $\mu$.  In the simulations we set $t=1$.  The factor $N$ is
introduced for convenience; with this definition, the large-$N$ limit
is defined by taking $N\to\infty$ keeping $\beta$ fixed~\cite{CR-93}.

One can easily check that Hamiltonian (\ref{hcpnla}) is invariant
under the global and local U(1) transformations (\ref{unsym}) and
(\ref{u1gausym}).  Moreover, by integrating out the gauge field, one
can rewrite the partition function as
\begin{equation}
Z = \sum_{\{z,\lambda\}} e^{-\beta H_{\lambda}} = \sum_{\{z\}}
\prod_{{\bm x},\mu} 
I_0\left(2\beta N t|\bar{\bm z}_{\bm x} \cdot {\bm z}_{{\bm
    x}+\hat\mu}|\right),
\label{partzint}
\end{equation}
where $I_0(x)$ is a modified Bessel function.  The corresponding
effective Hamiltonian is
\begin{equation}
H_{\lambda,{\rm eff}} = - \beta^{-1} 
\sum_{{\bm x},\mu}  
\ln I_0\left(2\beta N t|\bar{\bm{z}}_{\bm x} \cdot {\bm
      z}_{{\bm x}+\hat\mu}|\right).
\label{hlaeff}
\end{equation}
We recall that $I_0(x) = I_0(-x)$, $I_0(x) = 1 + x^2/4 + O(x^4)$, and
$I_0(x)\approx e^x/\sqrt{2\pi x}$ for large $x$.  The above formulas
imply that, independently of the sign of $t$, the model defined by the
Hamiltonian (\ref{hlaeff}), or equivalently Hamiltonian
(\ref{hcpnla}), is ferromagnetic, as is the formulation (\ref{hcpn})
with $J>0$. For $N=2$ Hamiltonian (\ref{hlaeff}) is a variant of the
standard O(3)-symmetric (Heisenberg) spin model, which is equivalent
to formulation (\ref{hcpn}).

The lattice formulation (\ref{hcpnla}) is numerically more convenient
than the formulation (\ref{hcpn}). The main reason is that Hamiltonian
(\ref{hcpnla}) is linear with respect to all variables ${\bm z}_{\bm
  x}$ and $\lambda_{{\bm x},\mu}$, unlike the standard Hamiltonian
(\ref{hcpn}).  This leads to notable advantages for Monte Carlo (MC)
simulations. For linear Hamiltonians one can use overrelaxed
algorithms~\cite{CRV-92,DMV-04,Hasenbusch-17}, which are more
efficient than the standard Metropolis algorithm, the only one that
can be straightforwardly used for the nonlinear Hamiltonian
(\ref{hcpn}).  We employ overrelaxed updates obtained by a stochastic
mixing of microcanonical and standard Metropolis updates of the
lattice variables~\cite{footnoteMC}.

In our numerical study we consider cubic lattices of linear size $L$
with periodic boundary conditions.  We compute the energy density and
the specific heat, defined respectively as
\begin{eqnarray}
E = {1\over N V} \langle H_{\lambda} \rangle,\qquad
C ={1\over N^2 V}
\left( \langle H_{\lambda}^2 \rangle 
- \langle H_{\lambda} \rangle^2\right),
\label{ecvdef}
\end{eqnarray}
where $V=L^3$.  We consider correlations of the gauge invariant
operator $Q_{\bm x}^{ab}$ defined in Eq.~(\ref{qdef}).  Its two-point
correlation function is defined as
\begin{equation}
G({\bm x}-{\bm y}) = \langle {\rm Tr}\, Q_{\bm x}
Q_{\bm y} \rangle,  
\label{gxyp}
\end{equation}
where the translation invariance of the system has been taken into
account.  The susceptibility and the correlation length are defined as
\begin{eqnarray}
&&\chi =  \sum_{{\bm x}} G({\bm x}) = 
\widetilde{G}({\bm 0}), 
\label{chisusc}\\
&&\xi^2 \equiv  {1\over 4 \sin^2 (\pi/L)}
{\widetilde{G}({\bm 0}) - \widetilde{G}({\bm p}_m)\over 
\widetilde{G}({\bm p}_m)},
\label{xidefpb}
\end{eqnarray}
where $\widetilde{G}({\bm p})=\sum_{{\bm x}} e^{i{\bm p}\cdot {\bm x}}
G({\bm x})$ is the Fourier transform of $G({\bm x})$, and ${\bm p}_m =
(2\pi/L,0,0)$ is the minimum nonzero lattice momentum.  We also
consider the Binder parameter
\begin{equation}
U = {\langle \mu_2^2\rangle \over \langle \mu_2 \rangle^2} , \qquad
\mu_2 = {1\over V^2} \sum_{{\bm x},{\bm y}} {\rm Tr}\, Q_{{\bm
    x}} Q_{\bm y} .
\label{binderdef}
\end{equation}

\section{Summmary of finite-size scaling theory}
\label{sumfss}

In this section we summarize the finite-size scaling (FSS) theory at
continuous and first-order transitions. It will be useful later to
distinguish the nature of the transition in 3D CP$^{N-1}$ models.

\subsection{Continuous transitions}
\label{conttra}

At continuous transitions the FSS limit is obtained by taking
$\beta\to \beta_c$ and $L\to\infty$ keeping
\begin{equation}
X \equiv (\beta-\beta_c)L^{1/\nu}
\label{Xdef}
\end{equation}
fixed, where $\beta_c$ is the inverse critical temperature and $\nu$
is the correlation-length exponent.  Any RG invariant quantity $R$,
such as $R_\xi\equiv \xi/L$ and $U$, is expected to asymptotically
behave as
\begin{eqnarray}
R(\beta,L) = f_R(X) + O(L^{-\omega}),
\label{rsca}
\end{eqnarray}
where $f_R(X)$ is a function, which is universal apart from a trivial
normalization of its argument and which only depends on the shape of
the lattice and on the boundary conditions. In particular, the
quantity $R^* \equiv f_R(0)$ is universal.  The approach to the
asymptotic behavior is controlled by the universal exponent
$\omega>0$, which is associated with the leading irrelevant RG
operator.

Assuming that the scaling function of a RG invariant quantity $R_1$ is
monotonic---this is generally the case for $R_\xi$ and therefore we
will usually set $R_1 = R_\xi$---we may also write
\begin{equation}
R_2(\beta,L) = F_R(R_1) + O(L^{-\omega}),
\label{r12sca}
\end{equation}
for any $R_2\neq R_1$, where $F_R(x)$ is a universal scaling function
as well. Eq.~(\ref{r12sca}) is particularly convenient, as it allows 
a direct check of universality, without the need of tuning any parameter.

In order to estimate the exponent $\eta$, one may analyze the FSS
behavior of the susceptibility. It scales as
\begin{equation}
\chi(\beta,L) \sim L^{2-\eta} \left[ f_\chi(X) + O(L^{-\omega})\right],
\label{chisca}
\end{equation}
or, equivalently, as
\begin{equation}
\chi(\beta,L) \sim L^{2-\eta} \left[ F_\chi(R_\xi) + O(L^{-\omega})\right].
\label{chisca2}
\end{equation}
The behavior of the specific heat at the transition is~\cite{PV-02}
\begin{equation}
C(\beta,L) \approx C_{\rm reg}(\beta) + L^{\alpha/\nu} f_C(X) ,
\label{cvsca}
\end{equation}
where $C_{\rm reg}(\beta)$ denotes the regular background, which is an
analytic function of $\beta$. This contribution is the dominant one
for $\alpha<0$, or, correspondingly, for $\nu>2/3$.  When $\alpha>0$,
we may also write the above equation as
\begin{equation}
C(\beta,L)\approx L^{\alpha/\nu} F_C(R_\xi) + O(L^{-\alpha/\nu},
L^{-\omega}).
\label{cvcont}
\end{equation}

\subsection{First-order transitions}
\label{fotra}

At first-order transitions the probability distributions of the energy
and of the magnetization are expected to show a double peak for large
values of $L$.  Therefore, two peaks in the distributions are often
taken as an indication of a first-order transition.  However, as
discussed, e.g., in Refs.~\cite{Ape-90,FMOU-90,Mc-90,Billoire-95} and
references therein, the observation of two maxima in the distribution
of the energy is not sufficient to conclude that the transition is a
first-order one.  For instance, in the two-dimensional Potts model
with $q=3$ and $q=4$~\cite{FMOU-90,Mc-90}, double-peak distributions
are observed, even if the transition is known to be continuous.
Analogously, in the 3D Ising model the distribution of the
magnetization has two maxima \cite{TB-00}.  In order to identify
definitely a first-order transition, it is necessary to perform a more
careful analysis of the large-$L$ scaling behavior of the
distributions or, equivalently, of the specific heat and the of Binder
cumulants ~\cite{CLB-86,VRSB-93,LK-91,CPPV-04,CNPV-14}.

If $E_+$ and $E_-$ are the values of the energy corresponding to the
two maxima of the energy-density distribution, the latent heat
$\Delta_h$ is given by $\Delta_h = E_+ - E_-$.  An alternative
estimate of the latent heat can be obtained from the FSS
of the specific heat $C$.  According to the standard
phenomenological theory~\cite{CLB-86} of first-order transitions, for
a lattice of size $L$ there exists a value $\beta_{{\rm max},C}(L)$ of
$\beta$ where $C$ takes its maximum value $C_{\rm max}(L)$. For large
volumes, we have
\begin{eqnarray}
&&C_{\rm max}(L) = V\left[ {1\over 4} \Delta_h^2 + O(1/V)\right],
\label{cmaxsc}\\
&&\beta_{{\rm max},C}(L)-\beta_c\approx c\,V^{-1} \label{betamax},
\end{eqnarray}
where $V=L^d$.

The Binder parameter $U$ can also be used to characterize a
first-order transition.  As discussed in Ref.~\cite{VRSB-93} (see also
Ref.~\cite{CPPV-04} for the extension to models with continuous
symmetries), the distribution of the order parameter is also expected
to show two peaks at $M_+$ and $M_-$, $M_- < M_+$, with $M_- \to 0$ as
$L\to \infty$ since there is no spontaneous magnetization in the
high-temperature phase.  As a consequence, the behavior of the Binder
parameter $U(\beta,L)$ at fixed $L$ must show a maximum $U_{\rm
  max}(L)$ at fixed $L$ (for sufficiently large $L$) at $\beta =
\beta_{{\rm max},U}(L) < \beta_c$ with
\begin{eqnarray}
&&U_{\rm max} = a\,V + O(1)\,,\label{umaxsca}\\ &&\beta_{{\rm
    max},U}(L) - \beta_c \approx b \,V^{-1}\,.
\label{bpeakU}
\end{eqnarray}
Note that FSS also holds at first-order
transitions~\cite{NN-75,FB-82,PF-83,CPPV-04}, although it is more
sensitive to the geometry and and to the nature of the boundary
conditions~\cite{CNPV-14}; for instance, FSS differs for boundary
conditions that favor or do not favor the different phases coexisting
at the transition~\cite{HPV-18,PRV-18}.  In the case of cubic systems
with periodic boundary conditions, FSS behavior is typically characterized by
an effective exponent $\nu=1/d=1/3$ (thus $\alpha/\nu=d=3$).  An
exponent $\nu$ larger than $1/d = 1/3$ indicates a continuous
transition.

\section{Numerical analysis of the 3D CP$^{1}$ model}
\label{ncp1}

\begin{figure}[tbp]
\includegraphics*[scale=\graphicscale]{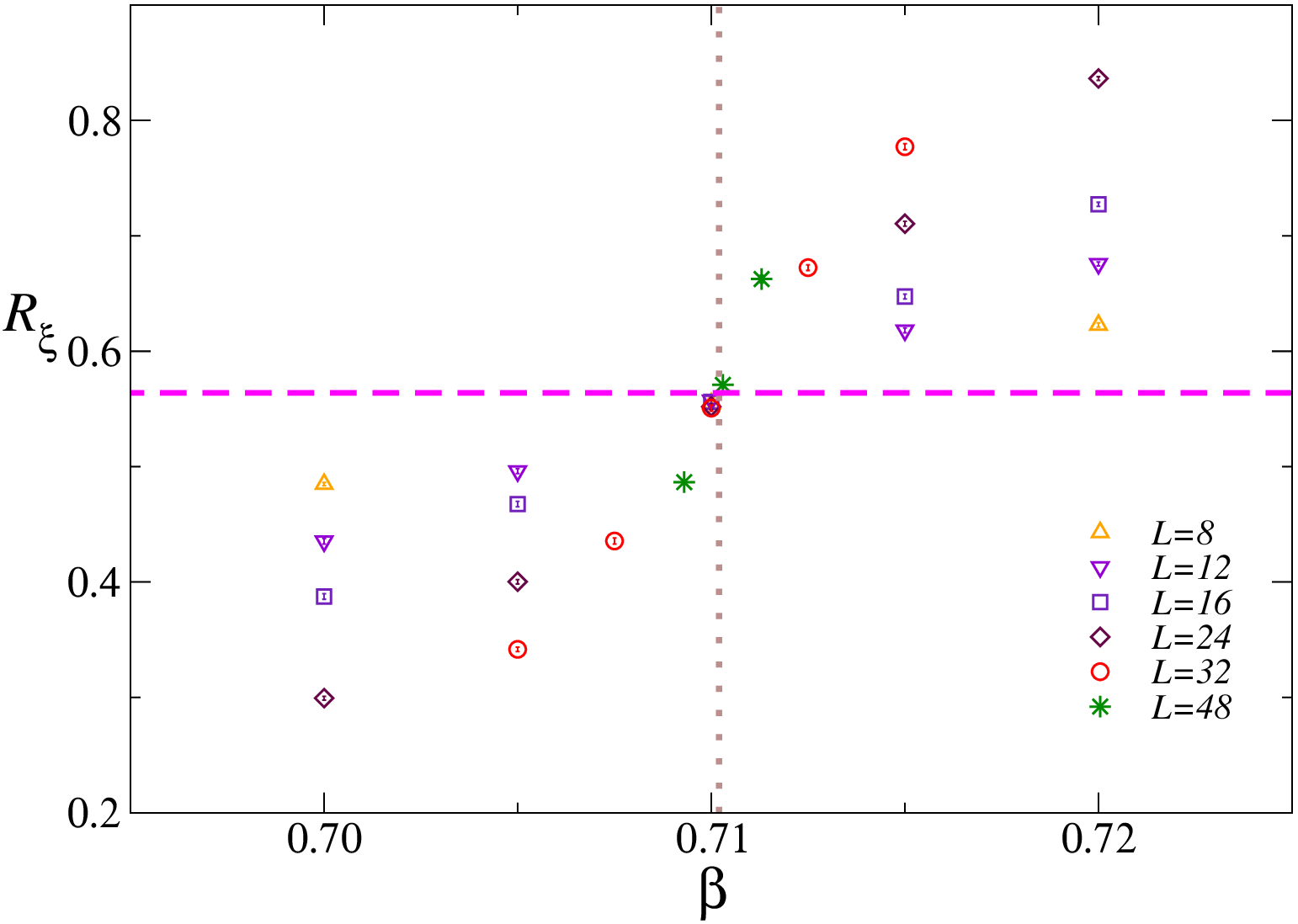}
\caption{Estimates of $R_\xi\equiv \xi/L$ for the CP$^1$ lattice model
  and several lattice sizes $L$ up to $L=48$.  The horizontal dashed
  line corresponds to the universal value $R_\xi^*=0.5639(2)$ of
  $R_\xi$ obtained for the 3D Heisenberg universality class.  The
  vertical dotted line indicates our best estimate $\beta_c=0.7102(1)$
  of the critical point.}
\label{rxin2}
\end{figure}

\begin{figure}[tbp]
\includegraphics*[scale=\graphicscale]{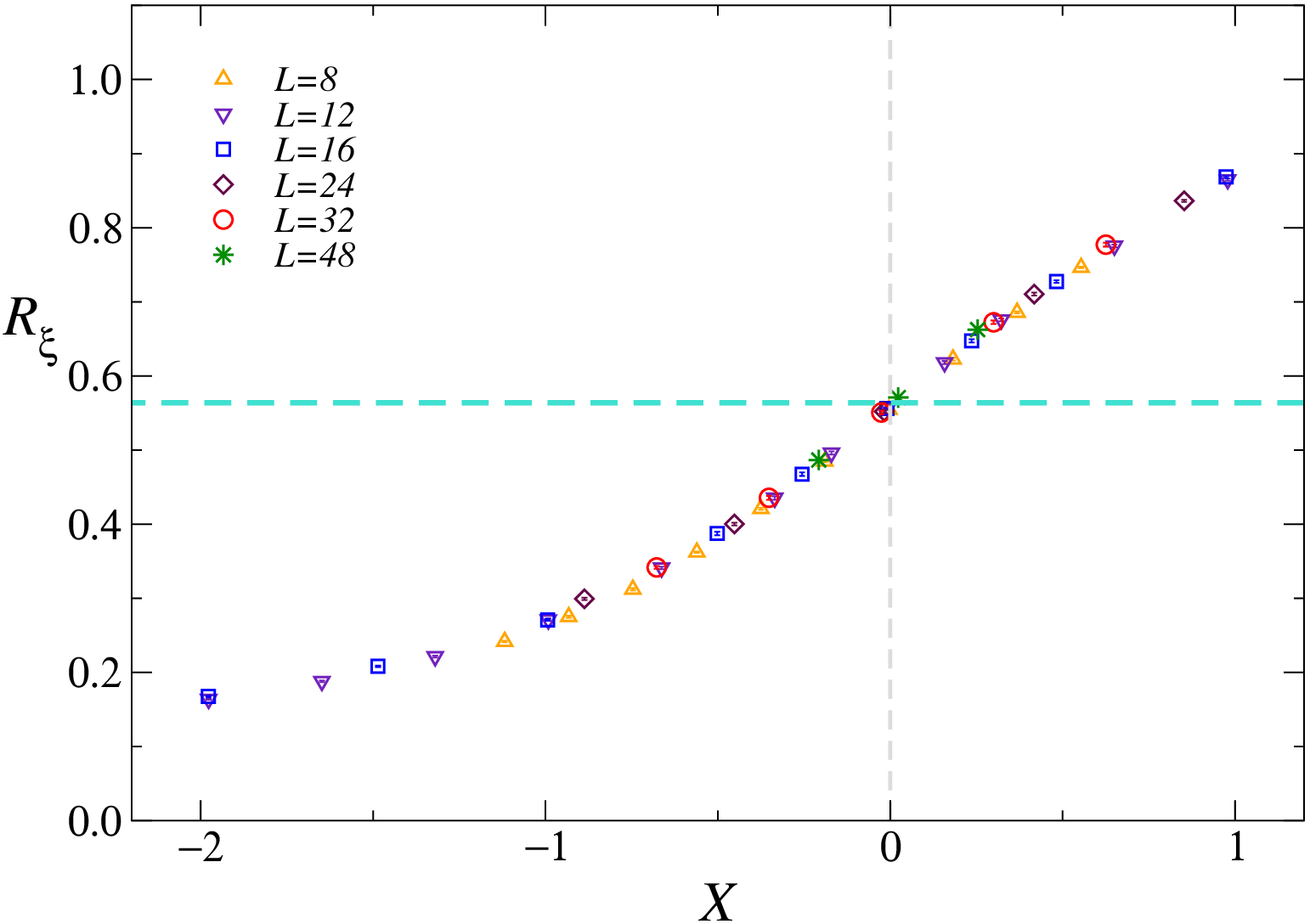}
\caption{Scaling plot of $R_\xi$ for the CP$^1$ model, versus $X
  \equiv (\beta - \beta_c)L^{1/\nu_h}$, using $\nu_{h}=0.7117$ and
  $\beta_c=0.7102$.  The horizontal dashed line corresponds to the
  universal value $R_\xi^*=0.5639(2)$ at the critical point ($X=0$)
  for the 3D Heisenberg universality class.  }
\label{scalrxin2}
\end{figure}

\begin{figure}[tbp]
\includegraphics*[scale=\graphicscale]{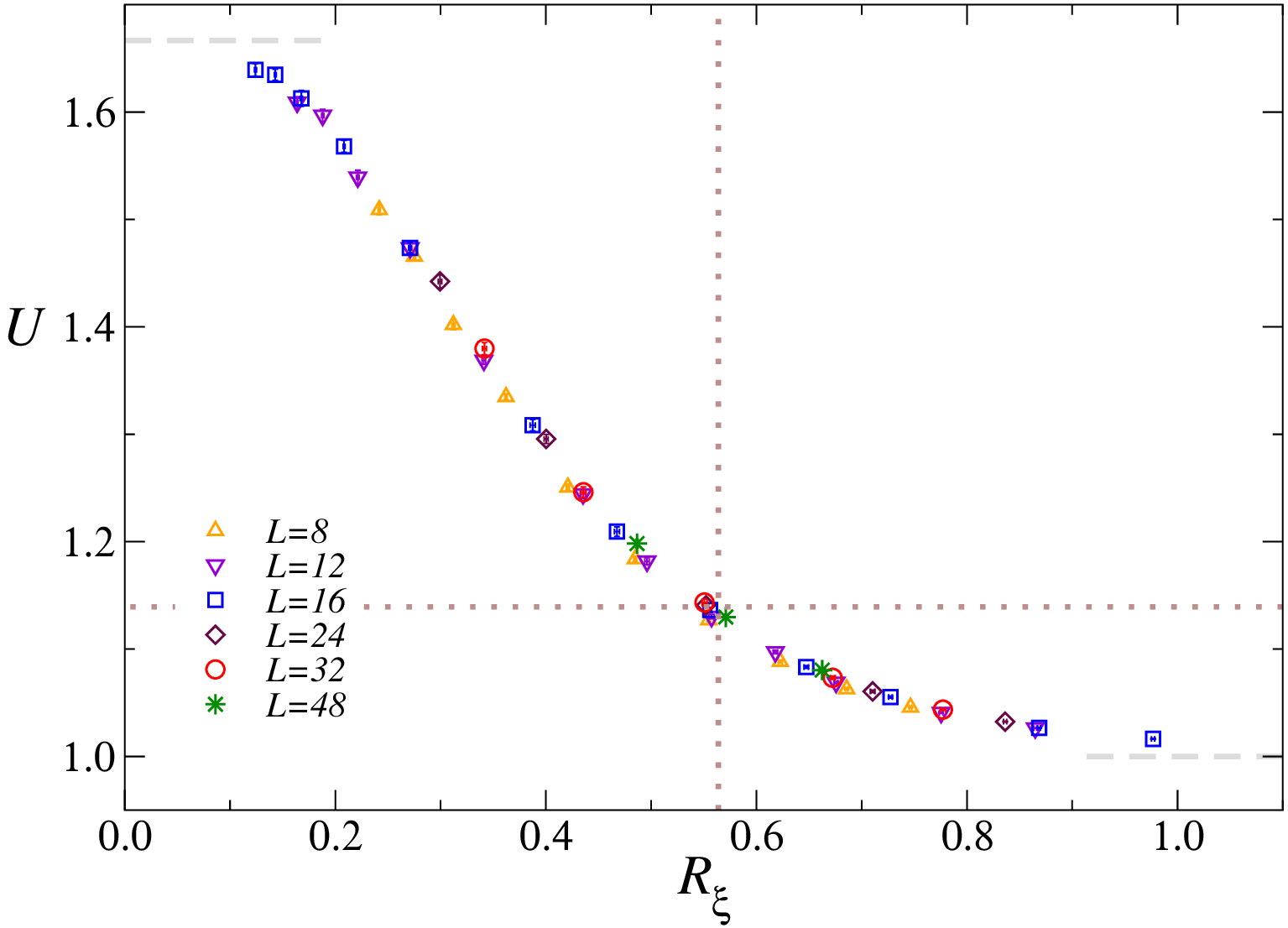}
\caption{Plot of $U$ versus $R_\xi$ for the CP$^1$ model.  The
  horizontal and vertical dotted lines correspond to the universal
  values $U^*=1.1394(3)$ and $R_\xi^*=0.5639(2)$, which are the
  universal values of $R_\xi$ and $U$ at the critical point for the 3D
  Heisenberg universality class, respectively. }
\label{scalbin2}
\end{figure}

\begin{figure}[tbp]
\includegraphics*[scale=\graphicscale]{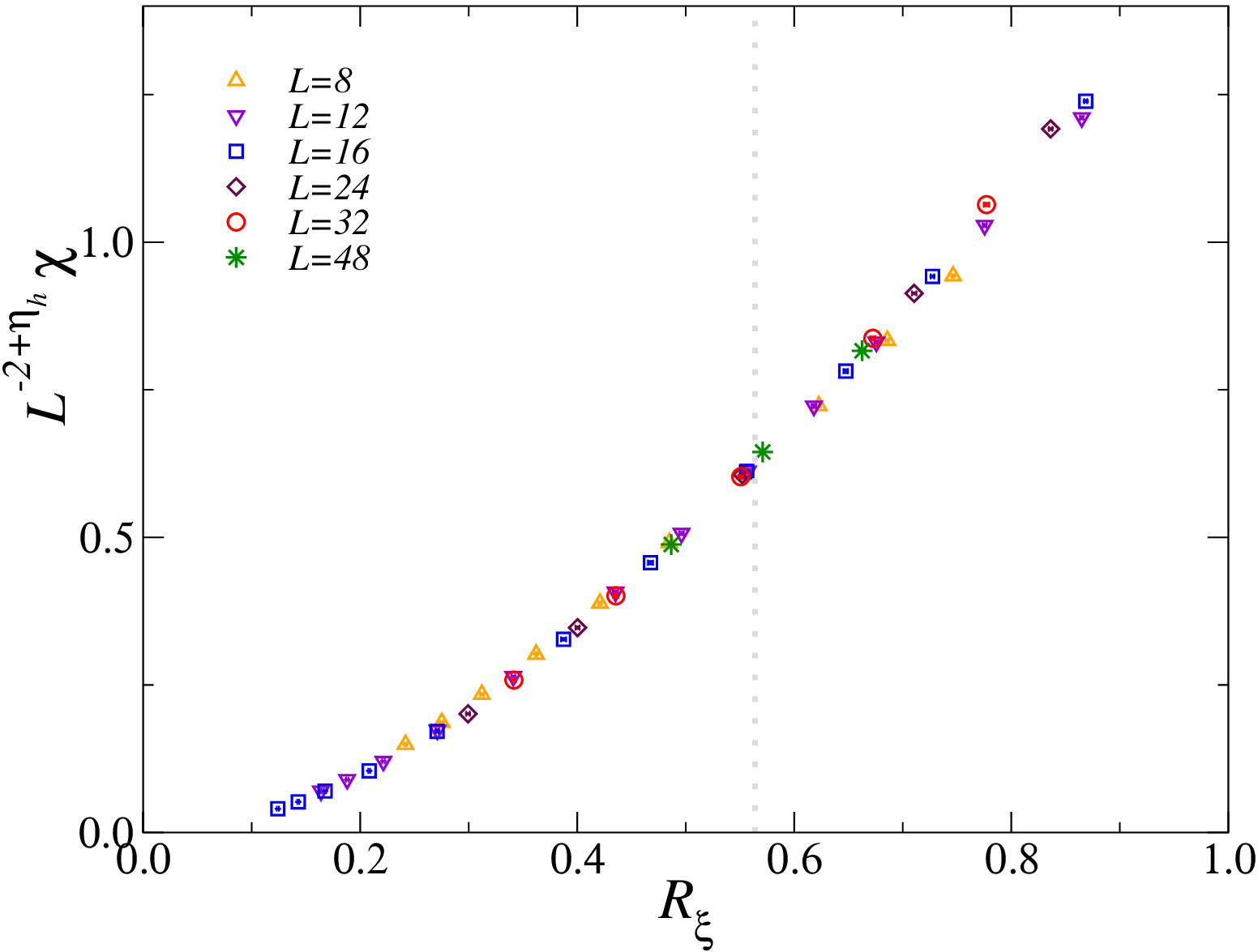}
\caption{ Scaling behavior of the susceptibility $\chi$ for the CP$^1$
  model: plot of $L^{-2+\eta_{h}} \chi$ versus $R_\xi$, using the 3D
  Heisenberg value $\eta_{h}=0.0378$.  }
\label{scalchin2}
\end{figure}

To begin with, we discuss the 3D CP$^1$ lattice model (\ref{hcpnla}),
corresponding to $N=2$.  As already mentioned in the introduction, the
transition in 3D CP$^1$ models should belong to the 3D O(3) vector
(Heisenberg) universality class, whose universal features are known
with high accuracy~\cite{PV-02}. Accurate estimates of the critical
exponents have been obtained by various methods; we
quote~\cite{HV-11,CHPRV-02}
\begin{eqnarray}
\nu_{h} = 0.7117(5),\qquad \eta_{h} =  0.0378(3),\label{o3exp}
\end{eqnarray}
so that  $\alpha_{h} = 2 - 3\nu_{h} = - 0.1351(15)$.  Moreover, accurate
results have also been obtained for the RG invariant quantities
$R_\xi$ and $U$. For cubic systems with periodic
boundary conditions, we have~\cite{HV-11} 
\begin{equation}
R_\xi^* =0.5639(2),\qquad U^*=1.1394(3),
\label{rxiustar}
\end{equation}
at the critical point.

In order to verify that the critical behavior of the lattice CP$^1$
model (\ref{hcpnla}) belongs to the 3D Heisenberg universality class,
we show that the FSS behavior in the CP$^{1}$ model has the same
universal features (same critical exponents and same values for
$R_\xi^*$ and $U^*$) as in the 3D Heisenberg model.

Figure \ref{rxin2} shows our results for the ratio $R_\xi=\xi/L$.  The
data corresponding to different value of $L$ have a crossing point,
which provides us a first estimate of the critical temperature,
$\beta_c\approx 0.710$.  The slopes of the curves are related to the
critical exponent $\nu$.  Their behavior is nicely consistent with the
Heisenberg value.  An accurate estimate of the critical point is
obtained by assuming the Heisenberg critical exponents and fitting the
data to Eq.~(\ref{rsca}).  In particular, we fit $R_\xi$ to the simple
ansatz
\begin{equation}
R_\xi = R_\xi^* + c_1\, X,\qquad X = (\beta - \beta_c)L^{1/\nu_{h}},
\label{rxisa}
\end{equation}
using the known estimates of $R_\xi^*$ and $\nu_{h}$. In the fit we
only consider the data belonging to a small interval around $\beta_c$,
where the behavior of $R_\xi$ looks linear.  We obtain $\beta_c =
0.7102(1)$.  Fig.~\ref{scalrxin2} shows $R_\xi$ versus $X \equiv
(\beta-\beta_c)L^{1/\nu_{h}}$.  The quality of the collapse of the
data onto a unique scaling curve provides a striking evidence that the
exponent $\nu$ for the lattice CP$^1$ model (\ref{hcpnla}) is that of
the Heisenberg universality class.  Scaling corrections are very
small; they are expected to decay as $L^{-\omega_{h}}$
with~\cite{GZ-98,PV-02,HV-11} $\omega_{h}=0.78(1)$.  We also note the
agreement of the $X=0$ value of the curves with the estimate
(\ref{rxiustar}) of $R_\xi^*$.

The Heisenberg nature of the transition is also supported by the data
for the Binder parameter $U$. In Fig.~\ref{scalbin2} we plot $U$
versus $R_\xi$.  The data scale on a single curve, in agreement with
Eq.~(\ref{r12sca}), and $U$ takes the value $U^*$ reported in
Eq.~(\ref{rxiustar}) when $R_\xi = R^*_\xi$.

The scaling behavior of the susceptibility is also in agreement with
Heisenberg behavior.  If we fix $\eta_h$ to the Heisenberg value, we
observe very good scaling, see Fig.~\ref{scalchin2}.  Since
$\alpha_{h}<0$, the leading asymptotic behavior of the specific heat
is given by the nonuniversal regular part, see
Eq.~(\ref{cvsca}). Thus, $C$ does not show any particular feature that
can be used to identify the universality class.

We may safely conclude that the above numerical FSS analysis confirms
that the critical behavior of the lattice formulation (\ref{hcpnla})
for $N=2$ belongs to the 3D O(3) vector universality class, as
expected.

\section{Numerical analysis of the 3D CP$^{3}$ model}
\label{ncp3}

In this section we discuss the CP$^3$ lattice model, i.e. model
(\ref{hcpn}) for $N=4$. This model wes already considered in
Ref.~\cite{KHI-11}, where the transition was identified as being of
first order.  Here, we present additional simulations on larger
lattices, which confirm the first-order nature of the
finite-temperature transition, in agreement with the predictions of
the corresponding LGW field theory.

\begin{figure}[tbp]
\includegraphics*[scale=\graphicscale]{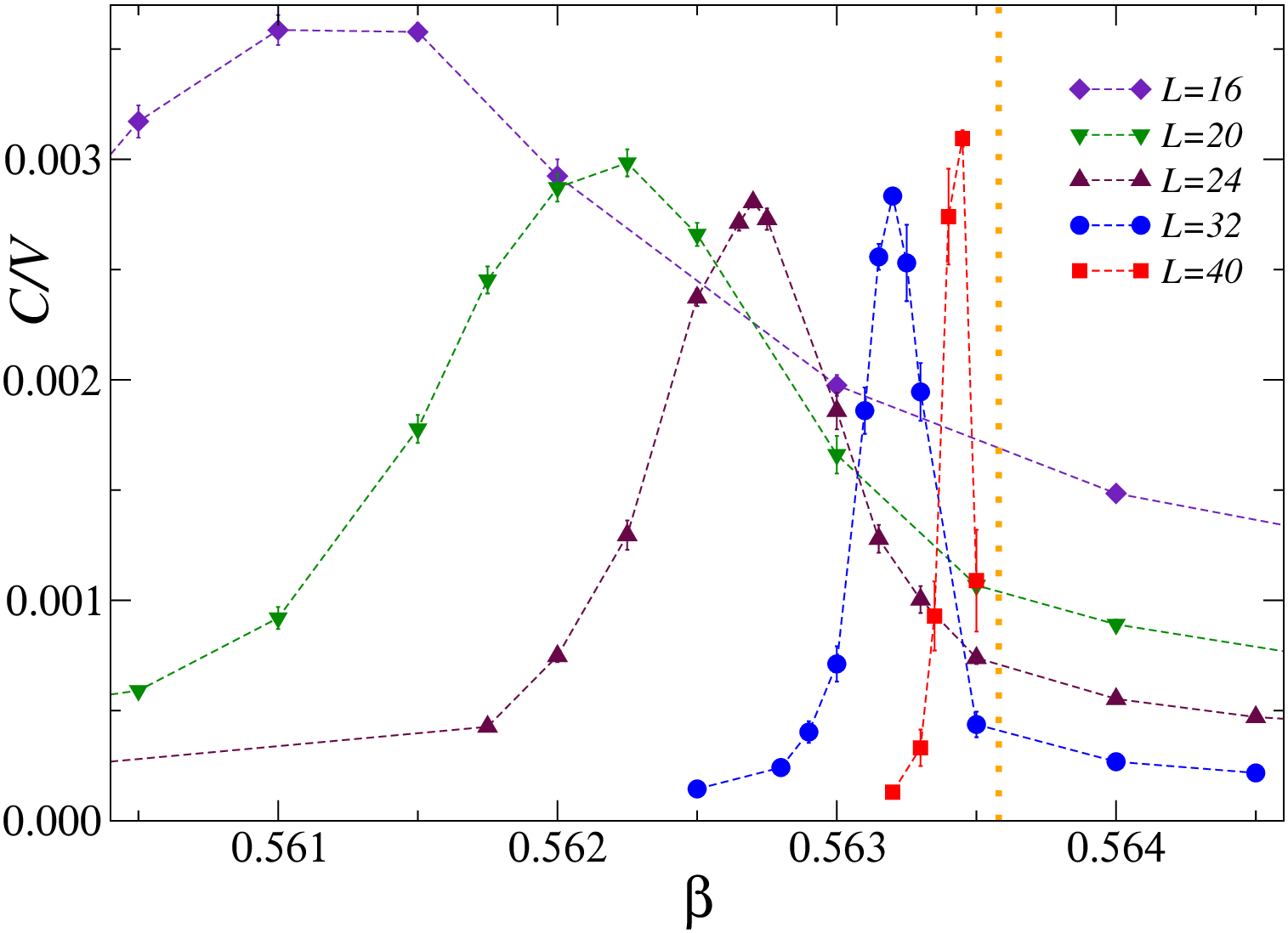}
\caption{Plot of the ratio $C/L^3$ versus $\beta$ for several lattice
  sizes $L$ up to $L=40$, where $C$ is the specific heat.  Results for
  the CP$^3$ lattice model.  The vertical dotted line correponds to
  the estimate $\beta_c\approx 0.5636$, obtained by extrapolating the
  position of the maximum of the specific heat, using
  Eq.~(\ref{betamax}).  The dotted lines connecting the data for the
  same size $L$ are only meant to guide the eye.  }
\label{cvovn4}
\end{figure}

\begin{figure}[tbp]
\includegraphics*[scale=\graphicscale]{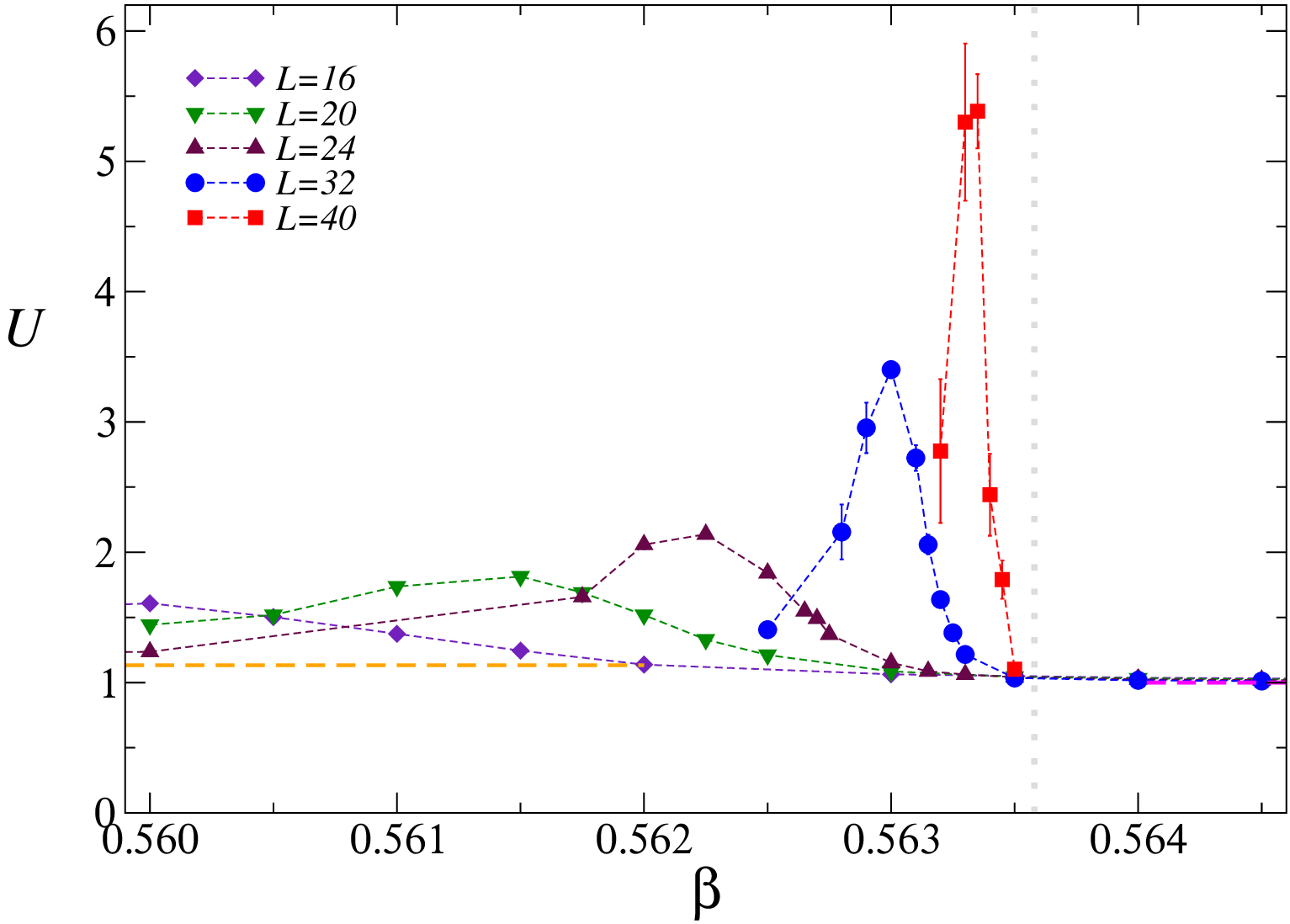}
\caption{Estimates of the Binder parameter $U$ for the CP$^3$ lattice
  model, for several lattice sizes $L$ up to $L=40$.  The vertical
  dotted line corresponds to the estimate $\beta_c\approx 0.5636$,
  while the horizontal dashed line corresponds to $U_h=17/15$, the
  high-temperature value of $U$.  The dotted lines connecting the data
  for the same size $L$ are only meant to guide the eye.}
\label{bin4}
\end{figure}

\begin{figure}[tbp]
\includegraphics*[scale=\graphicscale]{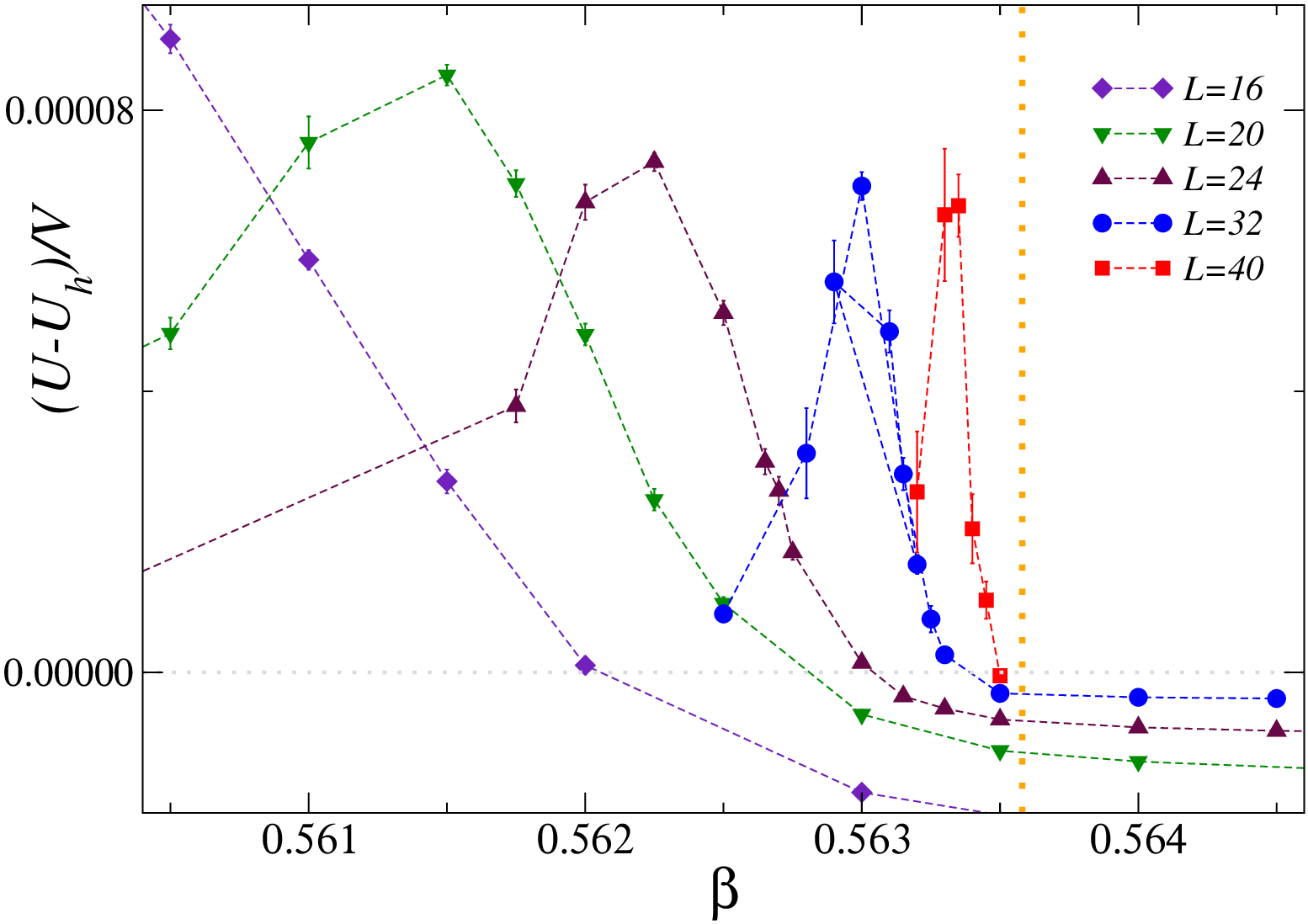}
\caption{Estimates of $(U-U_h)/L^3$ in the CP$^3$ lattice model for
  several lattice sizes $L$ up to $L=40$.  Here $U$ is the Binder
  parameter and $U_h=17/15$ its high-temperature value.  The vertical
  dotted line corresponds to the estimate $\beta_c\approx 0.5636$.
  The dotted lines connecting the data for the same size $L$ are only
  meant to guide the eye. }
\label{biovn4}
\end{figure}

In Fig.~\ref{cvovn4} we report the specific heat up to $L=40$.  For
each $L$, it shows a maximum which increases approximately as the
volume, thus approaching the asymptotic behavior predicted by
Eq.~(\ref{cmaxsc}) for a first-order transition.  We extrapolate the
position $\beta_{{\rm max},C}$ of the maximum using
Eq,~(\ref{betamax}), obtaining $\beta_c = 0.5636(1)$ for the
transition point. Moreover, using Eq.~(\ref{cmaxsc}) we can also
estimate the latent heat, obtaining $\Delta_h=0.11(1)$.

The first-order nature of the transition is also supported by the
behavior of the Binder parameter $U$. For each $L$, data show a peak at
a value $\beta_{{\rm max},U}(L)$, where $U$ is significantly larger than
the high-temperature and low-temperature values given by
\begin{eqnarray}
&&U_h\equiv \lim_{\beta\to 0} U = {N^2+1\over N^2-1}=
{17\over 15}\,,\label{uhlcp3} \\
&&U_l\equiv \lim_{\beta\to \infty} U = 1\,.\nonumber
\end{eqnarray}
Data are in substantial agreement with the predictions (\ref{umaxsca})
and (\ref{bpeakU}). This is clearly shown in Fig.~\ref{biovn4}, where
we plot $(U-U_h)/L^3$ ($U_h=17/15$ is the high-temperature value). As
$L$ increases the maximum of such quantity apparently approaches a
finite nonzero value, in agreement with Eq.~(\ref{umaxsca}).  Note
that the subtraction of the constant $U_h$ does not change the
asymptotic $L^3$ behavior. It is natural to expect that it provides a
reasonable approximation of the $O(1)$ contribution in
Eq.~(\ref{umaxsca}) (we subtracted $U_h=17/15$ instead of $U_l=1$,
because the maximum is located in the high-temperature phase).
Moreover, the large-$L$ extrapolation of the position of the maximum
using Eq.~(\ref{bpeakU}) is consistent with the value obtained from
the analysis of the specific heat.

\begin{figure}[tbp]
\includegraphics*[scale=\graphicscale]{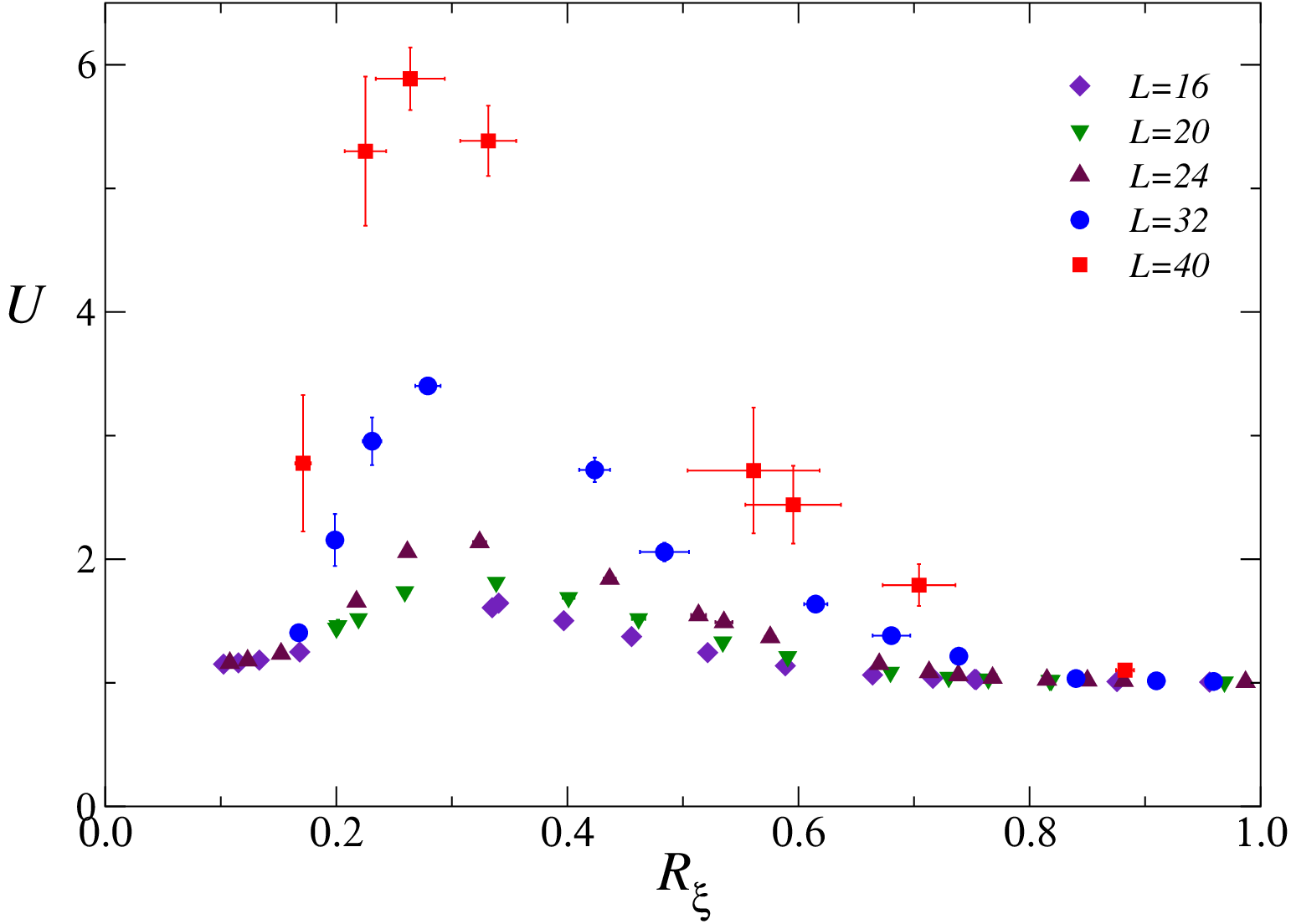}
\caption{ Plot of the Binder parameter $U$ versus $R_\xi$ for the
  CP$^3$ lattice model, for several lattice sizes $L$ up to $L=40$.
  Data do not scale, at variance with what happens in the CP$^1$ model,
 characterized by a continuous transition. }
\label{bin4rxi}
\end{figure}

Finally, in Fig.~\ref{bin4rxi} we plot $U$ versus $R_\xi$. The data do not
converge to an asymptotic curve, at variance with what happens in the CP$^1$
model where the transition is continuous, see Fig.~\ref{scalbin2}.

In conclusion, numerical results provide a robust evidence that the
finite-temperature phase transition is of first order in the CP$^3$
model (\ref{hcpn}).  This confirms the results of Ref.~\cite{KHI-11},
and, in particular, the standard LGW predictions. We recall
that a first-order transition was also found in
Refs.~\cite{NCSOS-11,NCSOS-13} for an alternative lattice loop
formulation corresponding to the 3D CP$^3$ model.

\section{Numerical analysis of the 3D CP$^{2}$ model}
\label{ncp2}

We now study the behavior of the 3D CP$^2$ model, which is more
controversial. On the one hand, the usual LGW scenario predicts a
first-order transition, as observed in the CP$^3$ model; on the other
hand, the finite-size scaling analysis \cite{NCSOS-11,NCSOS-13} of a
loop model that is expected to be in the same universality class as
that of the lattice CP$^2$ model favors a continuous transition with
critical exponents $\nu_n = 0.536(13)$, $\alpha_n = 0.39(4)$, and
$\eta_n=0.23(2)$. As we shall see, our FSS analysis of the lattice
CP$^2$ model up to $L=96$ favors a weak first-order transition.

\begin{figure}[tbp]
\includegraphics*[scale=\graphicscale]{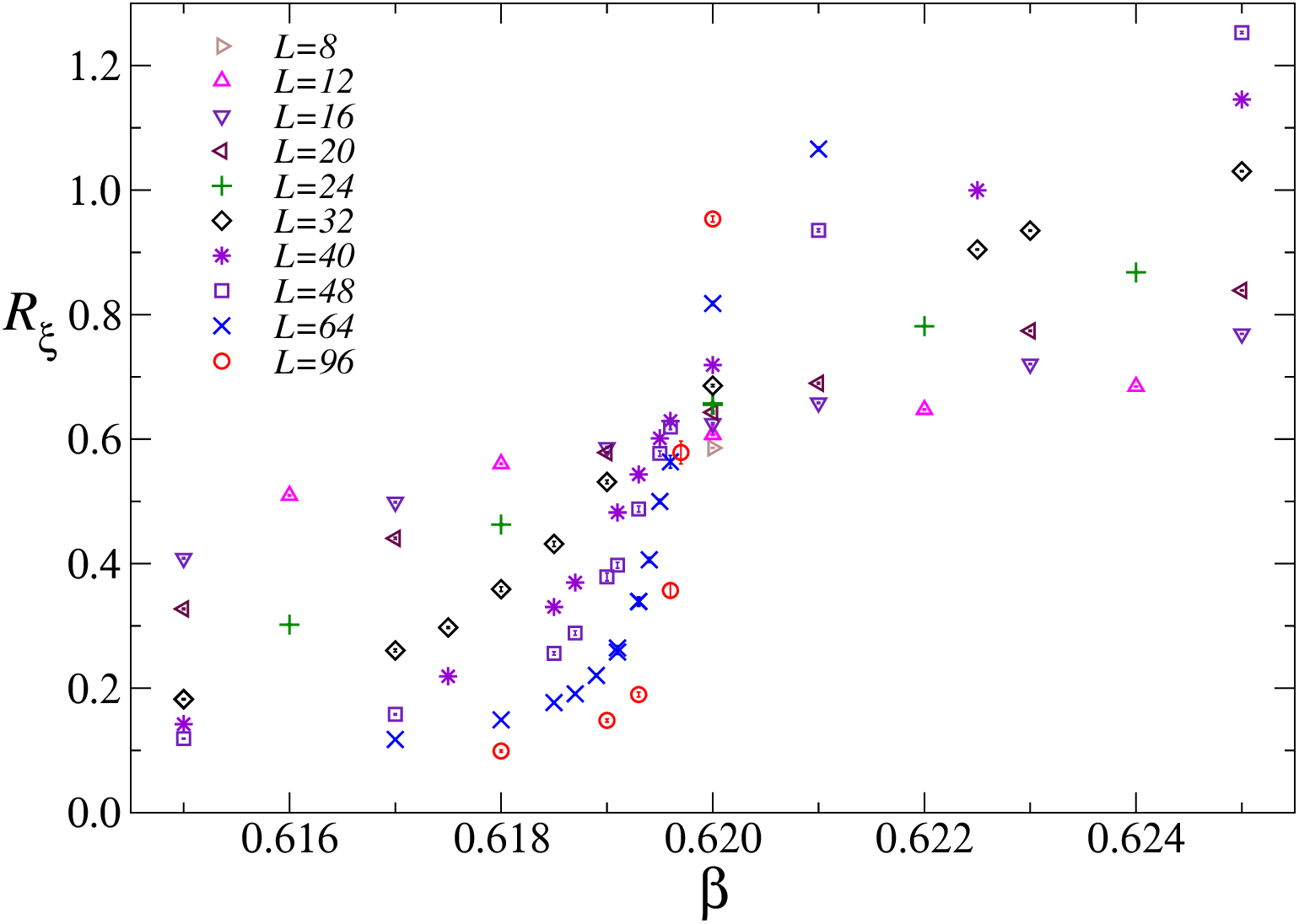}
\caption{Estimates of $R_\xi\equiv \xi/L$ for the CP$^2$ lattice model
  and several lattice sizes $L$ up to $L=96$.  }
\label{rxin3}
\end{figure}

\begin{figure}[tbp]
\includegraphics*[scale=\graphicscale]{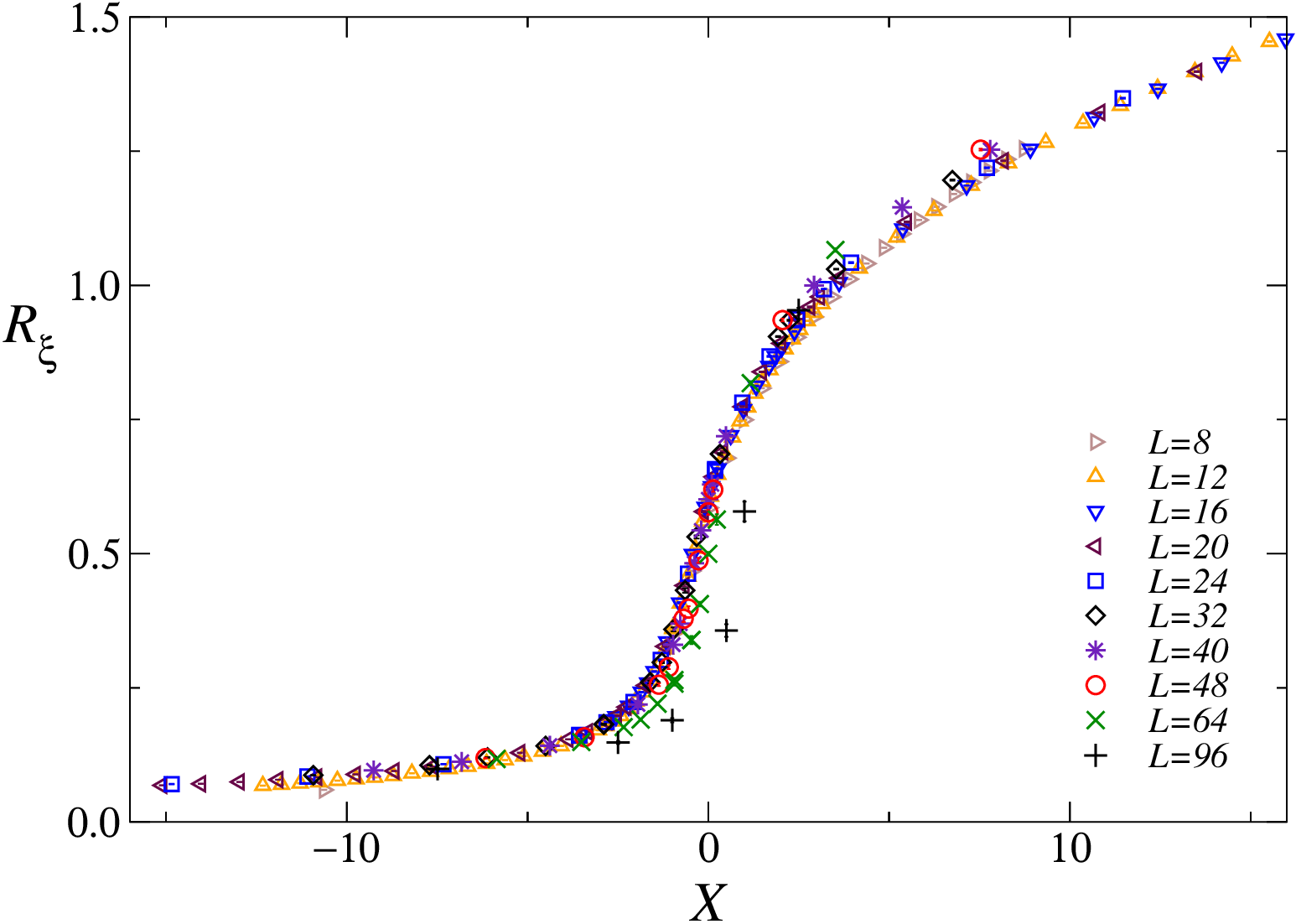}
\caption{Scaling plot of $R_\xi$ versus $X\equiv (\beta -
  \beta_c)L^{1/\nu_n}$, using $\nu_n=0.536$, which is the exponent
  obtained in Ref.~\cite{NCSOS-13}, and $\beta_c=0.6195$, which is
  obtained by fitting the data around $\beta_c$ (up to $L=48$) to
  Eq.~(\ref{rxisa}), keeping $\nu_n$ fixed.  Results for the CP$^2$
  model.  Data up to $L=48$ apparently collapse onto a single curve,
  but discrepancies appear when considering $L=64$ and $L=96$ data; no
  significant improvement is observed by only shifting the value of
  $\beta_c$.  }
\label{scalrxi}
\end{figure}

\begin{figure}[tbp]
\includegraphics*[scale=\graphicscale]{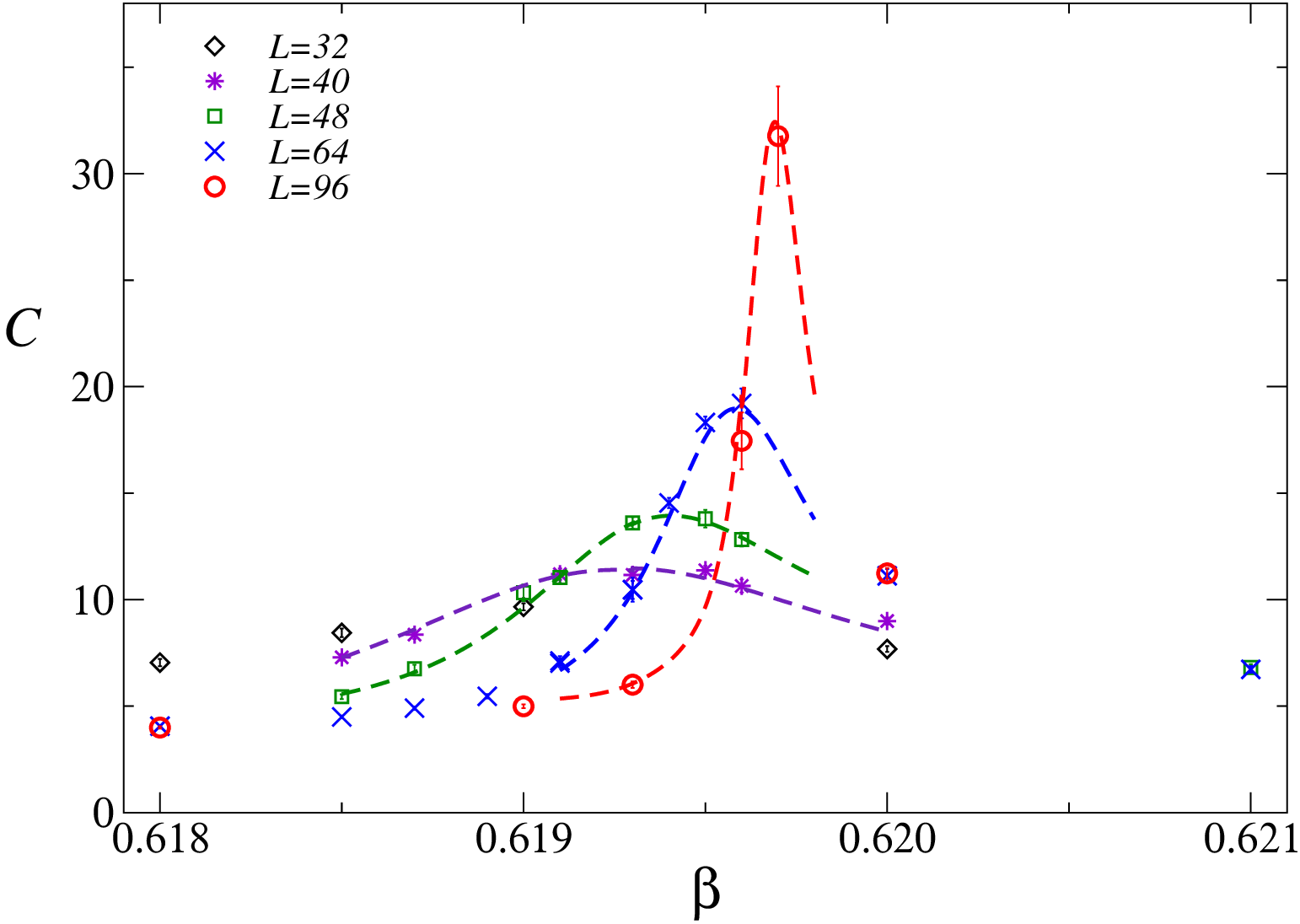}
\caption{Specific heat $C$ for the CP$^2$ lattice model as a function
  of $\beta$, for several lattice sizes $L$, up to $L=96$.  The dashed
  lines are interpolations obtained using reweighting tecniques
  \cite{FS-89}.  }
\label{cvn3}
\end{figure}

\begin{figure}[tbp]
\includegraphics*[scale=\graphicscale]{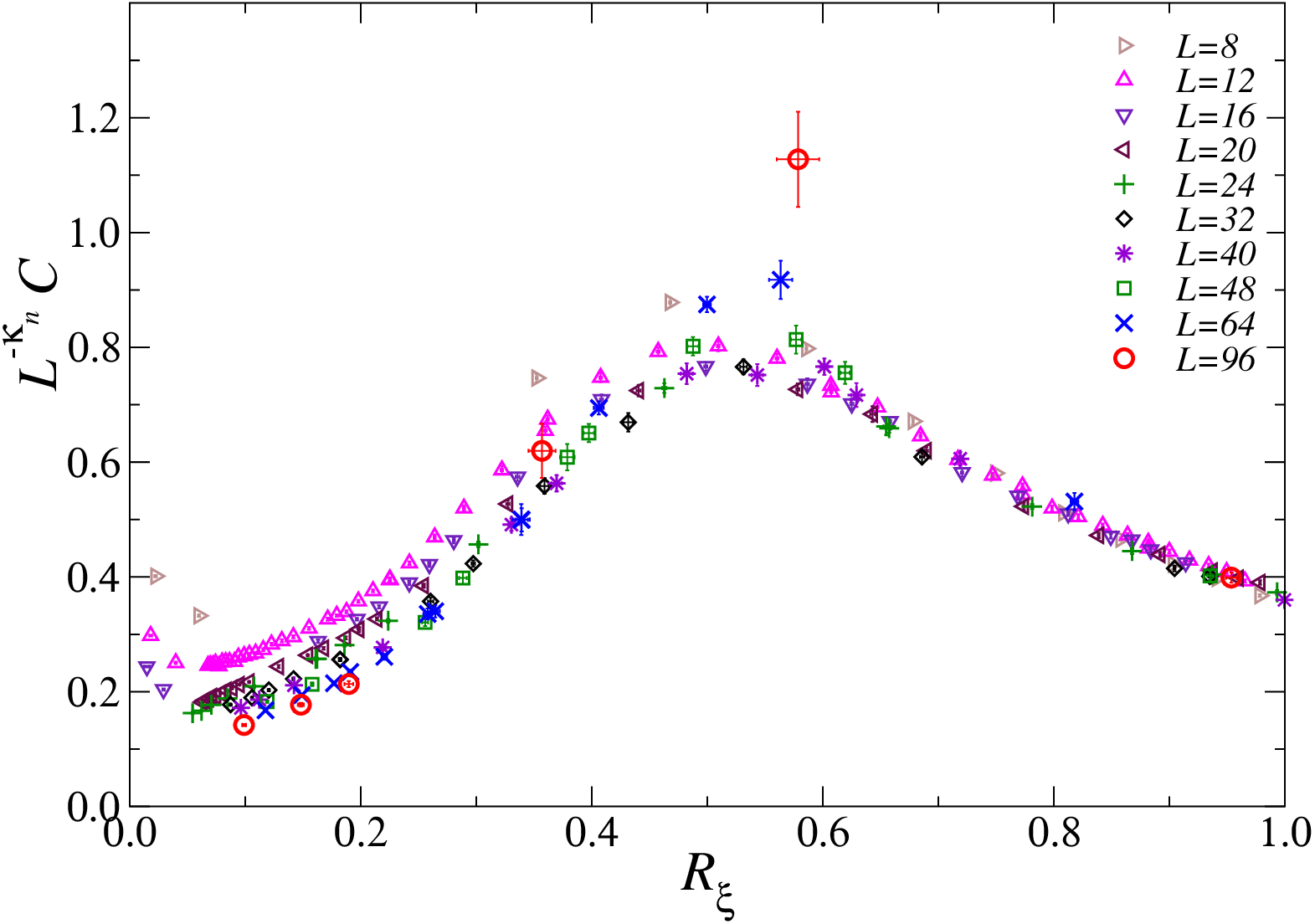}
\caption{ Plot of $L^{-\kappa_n} C$ versus $R_\xi$ for the CP$^2$
  lattice model.  We set $\kappa_n=0.73$, obtained using the relation
  $\kappa_n\equiv \alpha_n/\nu_n = (2/\nu_n - 3)$ and the estimate
  $\nu_n=0.536(13)$ of Ref.~\cite{NCSOS-13}. For a continuous
  transition with $\alpha>0$, data should asymptotically collapse onto
  a single curve, see Eq.~(\ref{cvcont}).  }
\label{scalcv}
\end{figure}

To identify the transition temperature, we consider the ratio $R_\xi$
as a function of $\beta$. As shown in Figure~\ref{rxin3}, the data for
different values of $L$ clearly show a crossing point, indicating a
finite-temperature transition for $\beta\approx 0.620$. If the CP$^2$
model has a continuous transition, $R_\xi(\beta,L)$ should scale as
predicted by Eq.~(\ref{rsca}). In this case, we expect the transition
to be in the same universality class as that discussed in
Ref.~\cite{NCSOS-13}, and hence exponents should be the same for the
two models. Therefore, we have analyzed our data setting $\nu =
0.536$, the estimate of Ref.~\cite{NCSOS-13}, and keeping $\beta_c$ as
a free parameter. We obtain a reasonable collapse of the data
corresponding to sizes $L\le 48$. Systematic deviations are instead
observed for the data with $L=96$ and, to a lesser extent, with
$L=64$, suggesting the presence of a crossover for $L\approx 100$. To
provide a more convincing evidence that the behavior for $L\le 48$ is
only a transient small-size behavior, we have repeated the
determination of $\beta_c$, using only the data corresponding to $L\le
48$.  We obtain $\beta_c \approx 0.6195$. The corresponding scaling
plot is reported in Fig.~\ref{scalrxi}. Up to $L=48$, the scaling is
good, but the largest-size data obtained on lattices with $L=64,96$
are clearly not consistent with a continuous transition with $\nu =
0.536$. We have also tested if the data for the susceptibility $\chi$
defined in Eq.~(\ref{chisusc}) satisfy the scaling behavior
(\ref{chisca2}), using the estimate $\eta_n=0.23(2)$ of
Ref.~\cite{NCSOS-13} ($\eta$ is the only free parameter). Scaling is
quite poor, even if we only consider data with $L\le 48$. For this set
of small-size results, a reasonable scaling is observed if we take a
significantly smaller value of $\eta$, $\eta\lesssim 0.1$. However,
the results for $L=96$ appear to be off the curve even for this value
of $\eta$.

Similar conclusions are reached from the analysis of the specific heat
$C$, reported in Fig.~\ref{cvn3}. In Fig.~\ref{scalcv}, we plot
$L^{-\kappa_n} C$ as a function of $R_\xi$, where $\kappa_n =
\alpha_n/\nu_n = (3/\nu_n - 1)$. Using \cite{NCSOS-13} $\nu_n =
0.536$, we fix $\kappa_n \approx 0.73$.  For a continuous transition,
all data should asymptotically fall onto a single curve, see
Eq.~(\ref{cvcont}).  The data up to $L=48$ are in reasonable agreement
with the predicted scaling behavior, but this is not the case for
those corresponding to $L=64$ and $96$.  The maximum $C_{\rm max}(L)$,
which can be computed quite precisely using the reweighting method of
Ref.~\cite{FS-89}, increases indeed faster than $L^{0.73}$, although
we are still very far from observing the asymptotic behavior $C_{\rm
  max}(L)\sim L^3$ appropriate for first-order transitions.

\begin{figure}[tbp]
\includegraphics*[scale=\graphicscale]{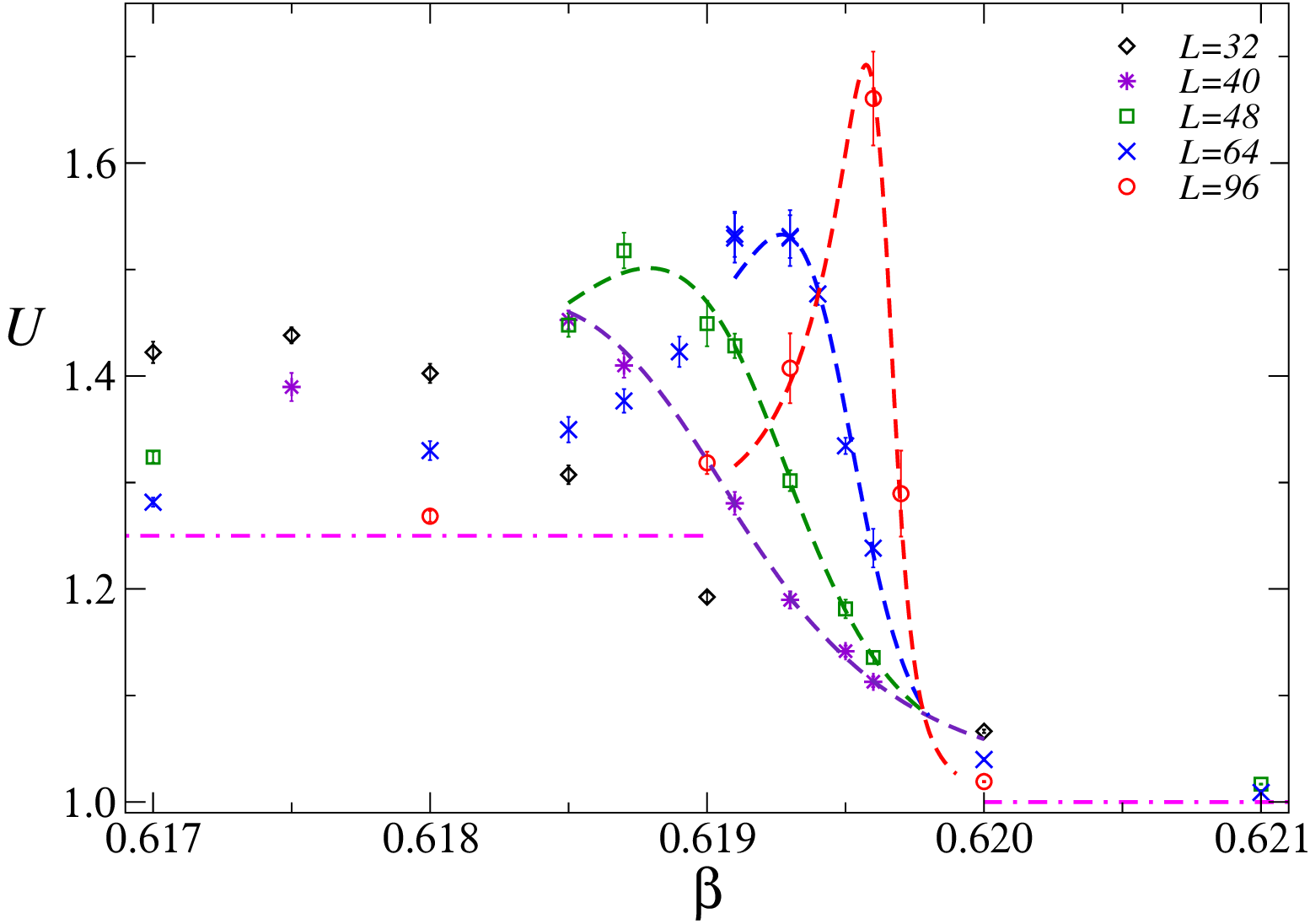}
\caption{Estimates of the Binder parameter $U$ for the CP$^2$ lattice
  model and several lattice sizes $L$ up to $L=96$.  The horizontal
  dashed lines correspond to $U=5/4$ and $U=1$, which are the
  asymptotic values of $U$ for $\beta\to 0$ and $\beta\to\infty$,
  respectively.  The interpolating dashed lines have been obtained
  using reweighting tecniques \cite{FS-89}.  }
\label{bin3}
\end{figure}

\begin{figure}[tbp]
\includegraphics*[scale=\graphicscale]{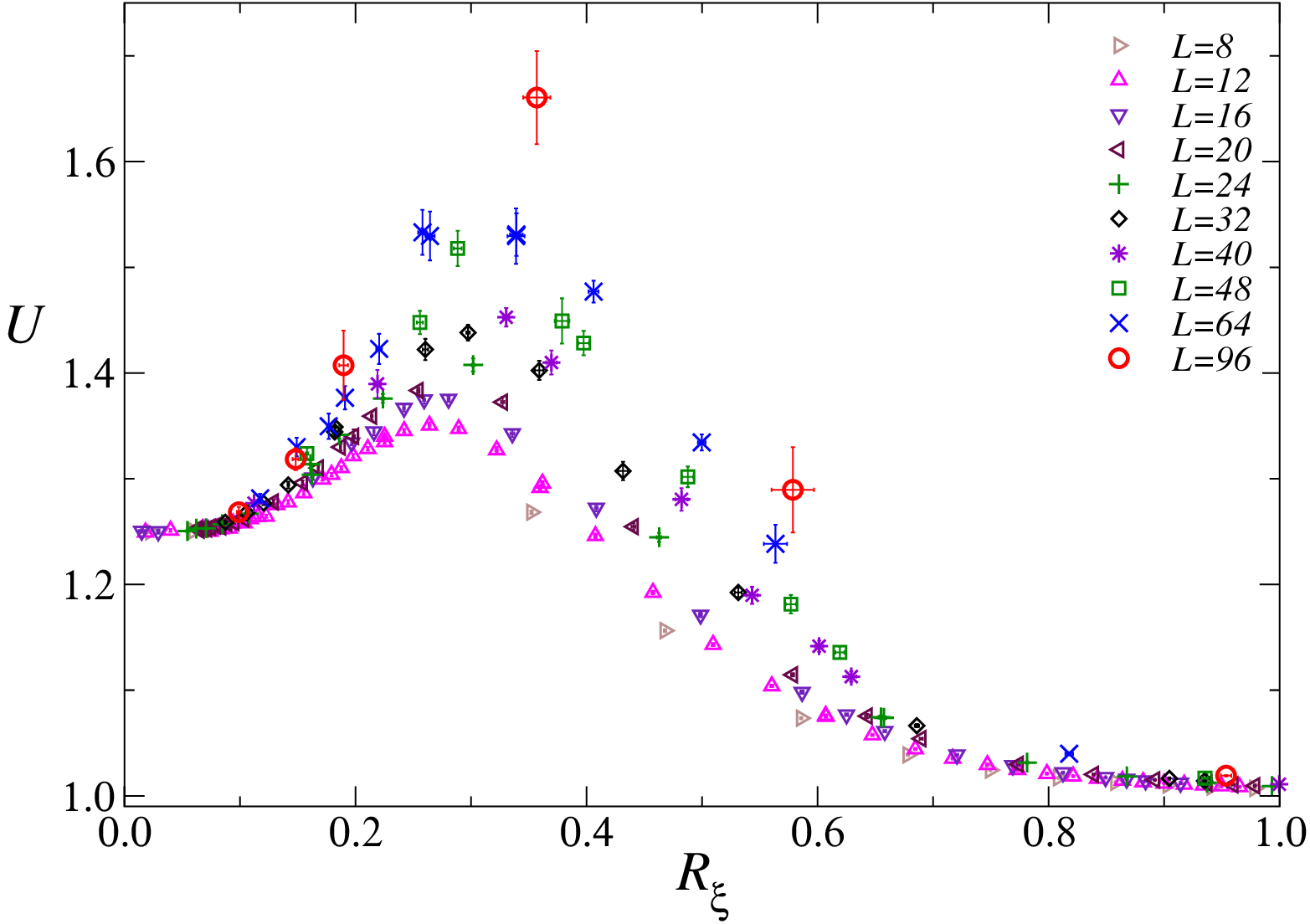}
\caption{The Binder parameter $U$ versus $R_\xi$ for the CP$^2$ model
  and several lattice sizes $L$ up to $L=96$.  }
\label{bine-rxi}
\end{figure}

The strongest indication for a first-order transition comes from the
anomalous behavior of the Binder parameter $U$. The numerical results,
reported in Fig.~\ref{bin3}, show that the maximum $U_{\rm max}(L)$
does not converge to a finite value as $L$ increases, at variance with
what is expected for a continuous transition. It is also useful to
plot $U$ versus $R_\xi$. For a continuous transition data should
collapse onto a single curve (note that there are no free parameters),
and indeed they do in the CP$^1$ model, see Fig.~\ref{scalbin2}.
Instead, here no evidence of scaling emerges, see Fig.~\ref{bine-rxi}.
The plot is similar to that obtained for the CP$^3$ model, see
Fig.~\ref{bin4rxi}, where the transition is of first order. Note that
the plot of $U$ versus $R_\xi$ is apparently the most appropriate one
to identify the nature of the transition. Indeed, at variance with the
previous analyses, here there are no transient deceiving effects: even
if we only consider data with, say, $L\le 32$, we would still observe
a poor scaling behavior in the interval $0.3 \lesssim R_\xi \lesssim
0.6$.

\begin{figure}[tbp]
\includegraphics*[scale=\graphicscale]{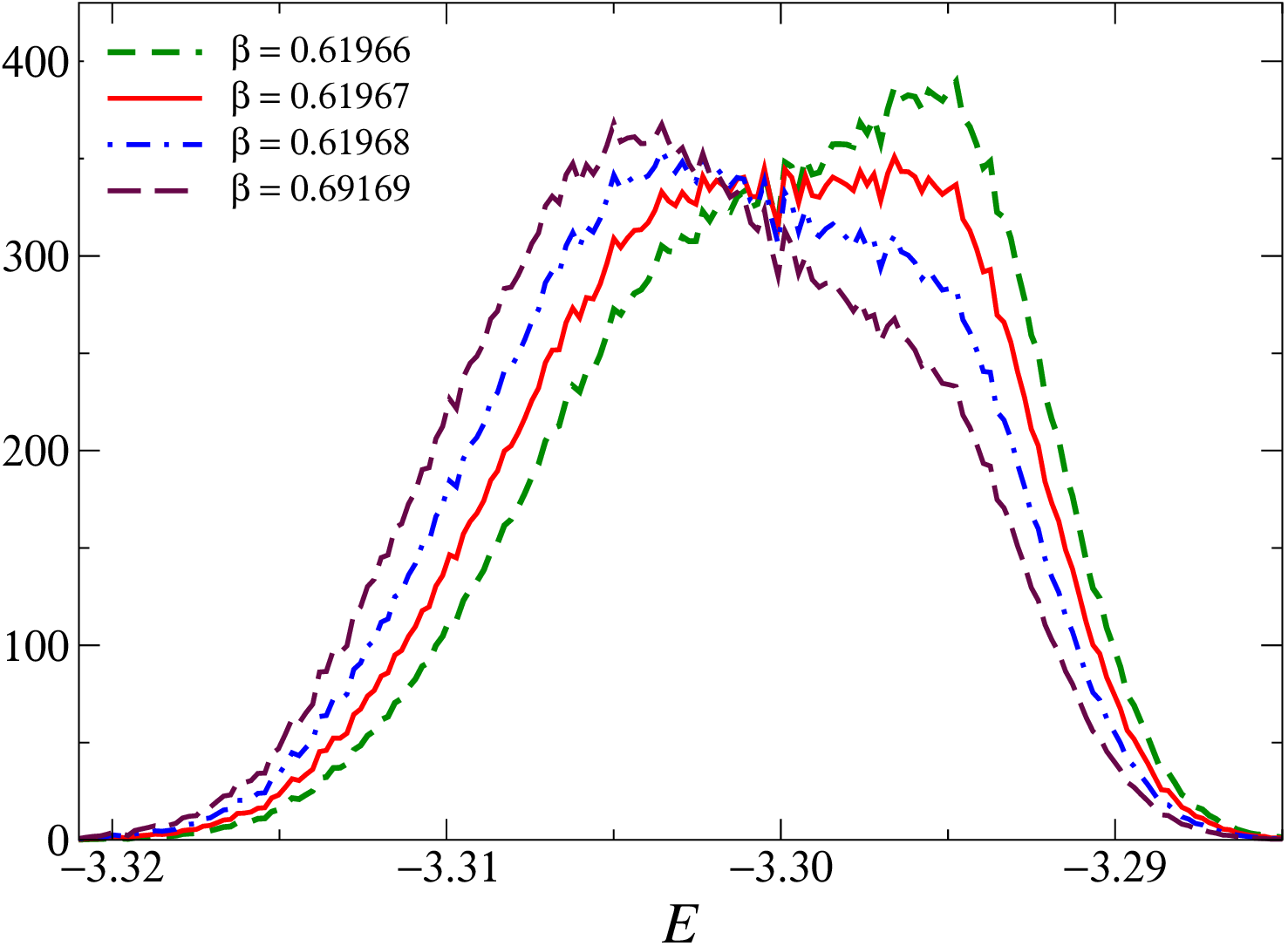}
\includegraphics*[scale=\graphicscale]{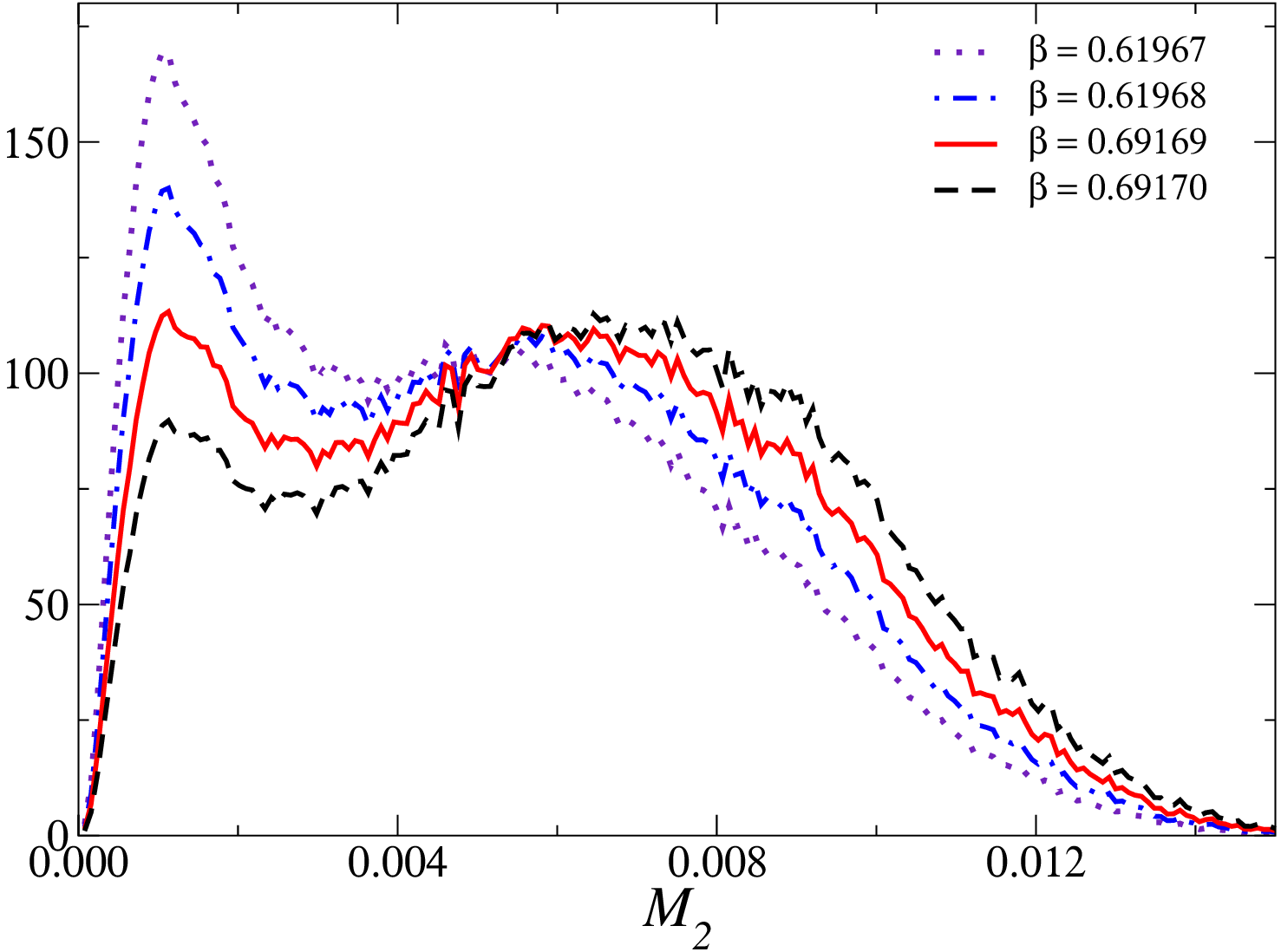}
\caption{Distributions of the energy density (top) and of $\mu_2$
  defined in Eq.~(\ref{binderdef}) (bottom); see
  Eq.~(\ref{distributions}).  We report results for several values of
  $\beta$ close to the transition point and $L=96$. The distributions
  have been estimated using reweighting techniques \cite{FS-89}.  }
\label{histos}
\end{figure}

Finally, we consider the distributions of the energy density and 
of the quantity $\mu_2$, defined in  Eq.~(\ref{binderdef}):
\begin{eqnarray}
&& P_E(E) = \langle \delta [E - H_\lambda/(NV)]\rangle, \qquad \nonumber \\
&& P_M(M_2) = \langle \delta (M_2 - \mu_2)\rangle .
\label{distributions}
\end{eqnarray}
Up to $L=64$, the distribution $P_M(M_2)$ does not show a double-peak
structure, although, for $L = 48$ and $L=64$, one can identify a
somewhat flat top region, which is analogous to that observed for the
energy when $L=96$ (see below). Two distinct peaks are only identified
for $L=96$, see Fig.~\ref{histos}.  In the case of the energy density
instead, even for $L=96$ we do not observe a double-peak
structure. However, the distribution corresponding to $\beta =
0.61967$ shows an intermediate flat top region, suggesting that two
peaks will appear for largest values of $L$.  The width of the flat
region (we make the reasonable assumption that such width is larger
than the distance between the positions of the two maxima that will
appear for large $L$) provides an upper bound for the latent heat. We
obtain $\Delta_h < 0.01$, which is very small, much smaller than
$\Delta_h\approx 0.11$ obtained for the CP$^3$ model.

In conclusion, the FSS analysis of the data up to $L\approx 100$
supports the presence of a weak first-order transition.  For
$L\lesssim 50$, most of the data show an intermediate regime where a
behavior compatible with a continuous transition is observed. In this
regime, results are roughly consistent with those obtained from the
analysis of the CP$^2$ loop model \cite{NCSOS-11,NCSOS-13} with system
sizes up to $L\approx 100$.  However, our data corresponding to $L=64,
96$ are not consistent with the conclusions of
Refs.~\cite{NCSOS-11,NCSOS-13}. It is interesting to note that the
most robust evidence against a continuous transition is provided by
the Binder parameter, which, however, was not considered in
Refs.~\cite{NCSOS-11,NCSOS-13}.

From our results we might conclude that no CP$^2$ universality class
exists.  In this case, the scaling
behavior observed in Ref.~\cite{NCSOS-13} is not asymptotic, but
simply an intermediate transient regime, as it occurs in the lattice
formulation we consider.  In this case, also in the loop model one
would observe first-order behavior for large values of $L$.  However,
we would like to stress that we cannot completely exclude that a
CP$^2$ fixed point really exists.  Indeed, it is {\em a priori}
possible that the CP$^2$ lattice model is outside the attraction
domain of the fixed point.  With increasing $L$, the
renormalization-group flow for the lattice model gets initially closer
to the CP$^2$ fixed point, giving somehow rise to an intermediate
scaling regime, and then it moves away to infinity, as expected for a
first-order transition.

\section{The lattice CP$^{N-1}$ models in the large-$N$ limit}
\label{ncpinf}

To obtain further insight on the critical behavior of CP$^{N-1}$
models, we consider the large-$N$ limit. We study a very general class
of Hamiltonians given by
\begin{equation}
H = - N \sum_{ {\bm x}\mu} 
    W (|\bar{\bm z}_{{\bm x}+\hat\mu} \cdot {\bm z}_{\bm x}|^2),
\label{generalH}
\end{equation}
where $W(x)$ is an increasing function of $x$ in $[0,1]$. 
Hamiltonian~(\ref{hcpn})
corresponds to $W(x) = x$. The gauge Hamiltonian~(\ref{hlg}) can also be studied
in this framework. Indeed, in the large-$N$ limit, 
Hamiltonian~(\ref{hlg}) is equivalent to 
\begin{equation}
H = - 2 N \sum_{ {\bm x}\mu} 
     |\bar{\bm z}_{{\bm x}+\hat\mu} \cdot {\bm z}_{\bm x}|.
\end{equation}
It therefore corresponds to the choice $W(x)= 2 \sqrt{x}$.

The general analysis for $N=\infty$ is presented in the Appendix.  We
find that a very general class of models undergoes a
finite-temperature first-order transition.  This class includes all
Hamiltonians for which $W'(x)$ is bounded in $[0,1]$. In particular,
it includes Hamiltonian (\ref{hcpn}), but not the gauge Hamiltonian
(\ref{hlg}) (since the latter corresponds to $W(x) = 2 \sqrt{x}$,
$W'(x) = 1/\sqrt{x}$ is not bounded in $[0,1]$) . This result
generalizes Ref.~\cite{Samuel-83}, that showed that model (\ref{hcpn})
has a first-order transition in two dimensions. If we additionally
assume that $W''(x) \ge 0$, we can completely characterize the
coexisting phases. In the high-temperature phase, the correlation
length vanishes identically and the system is disordered. In the
low-temperature phase, instead, the system is ordered and the global
symmetry is broken.  The behavior we find for large $N$ in this class
of systems is very similar to what we observe in the gauge model for
$N=3$ and 4.

It is also possible to observe continuous transitions. This is the
case of the gauge model (\ref{hlg}) or of linear combinations of the
Hamiltonians (\ref{hlg}) and (\ref{hcpn}).  It is important to stress
that the analysis shows that continuous transitions can be obtained
without needing an exact tuning of a Hamiltonian parameter. It
therefore confirms the existence of a stable CP$^{N-1}$ fixed point
for large $N$ \cite{MZ-03}.  However, this does not guarantee that all
CP$^{N-1}$ models undergo a continuous transition. There is indeed a
large class of theories that undergo a first-order transition.  In the
renormalization-group language, these models do not belong to the
attraction domain of the fixed point, so that, in the
renormalization-group evolution, they flow to infinity, giving rise to
a discontinuous nonuniversal transition.

\section{Summary and conclusions}
\label{conclu}

In the last fifty years the LGW approach has played a fundamental role
as it has provided both qualitative predictions and accurate
quantitative estimates for the universal properties of critical
transitions \cite{Fisher-75,Ma-book,PV-02}. The approach has also been
applied to systems with gauge symmetries, the most notable case being
QCD.  Considering an appropriate gauge-invariant observable as order
parameter, the LGW approach was used \cite{PW-84,BPV-03,PV-13} to
predict the nature of the chiral transition in the massless limit
of quarks.
that is presently looked for in heavy-ion collisions. However, we have
recently noted \cite{PTV-17,PTV-18} that, if a gauge invariant order
parameter is used, for some systems that enjoy a
gauge invariance the standard LGW scenario 
is not consistent with the numerical and/or
theoretical results obtained with other methods. This is the case of
ferromagnetic and antiferromagnetic CP$^{N-1}$ models and of
antiferromagnetic RP$^{N-1}$ systems \cite{PTV-17,PTV-18}, which are
invariant under U(1) and ${\mathbb Z}_2$ gauge transformations,
respectively.

Here we consider the ferromagnetic CP$^{N-1}$ models that are
invariant under U(1) transformations. If one adopts the usual LGW strategy,
one defines a gauge-invariant order
parameter and considers the most general LGW $\Phi^4$ theory
compatible with the symmetries, cf. Eq.~(\ref{hlg}).  In the case of
CP$^{N-1}$ models with $N\ge 3$, a cubic term is present in the LGW
Hamiltonian. In the mean-field approximation this implies that the transition
is of first-order. Such prediction
is expected to hold close to four dimensions,
where statistical fluctuations are not expected to be sufficiently
strong to change the nature of the transition, and, in particular, to be valid 
in three dimensions. In this scenario, one would predict 
the absence of continuous
transitions for all values of $N$ satisfying $N\ge 3$ in generic
CP$^{N-1}$ models.  If the model undergoes a finite-temperature
transition, as it usually does, such transition should generically be
of first order.

These conclusions disagree with the general arguments of
Ref.~\cite{MZ-03}, that discusses CP$^{N-1}$ models in the large-$N$
limit, and the numerical results of
Refs.~\cite{NCSOS-11,NCSOS-13} for $N=3$.   
In particular, for large values of
$N$, it si possible to show \cite{MZ-03} that the continuum CP$^{N-1}$
field theory is equivalent to an effective abelian Higgs model, for
which a large-$N$ stable fixed point can be identified using the
standard perturbative $\epsilon$ expansion \cite{HLM-74,MZ-03}.
Therefore, in this limit CP$^{N-1}$ models may undergo a continuous
finite-temperature transition.

In this work we reconsider the problem for ferromagnetic CP$^{N-1}$
models.  We perform numerical simulations of the lattice model
(\ref{hlg}), in which gauge invariance is guaranteed by the presence
of an explicit gauge U(1) link variable
\cite{RS-81,DHMNP-81,BL-81}. Such a formulation is particularly
convenient from a numerical point of view and allows us to obtain
accurate results for systems of size $L\le 96$. For $N=2$ we confirm
the equivalence of the CP$^{1}$ model with the O(3) Heisenberg vector
model, while, for $N=3$ and 4, we find that the transition is of first
order. The results for $N=4$ agree with those of Ref.~\cite{KHI-11},
obtained on smaller lattices.  For $N=3$ the first-order singularity
is very weak and indeed, a clear signature of the nature of the
transition is only obtained for sizes $L\gtrsim 50$. For smaller
sizes, most of the numerical data are consistent with a continuous
transition characterized by critical exponents roughly equal to those
obtained in Ref.~\cite{NCSOS-13}. Therefore, our numerical results for
$N=2$, 3, and 4 are consistent with the standard LGW picture, in which 
one predicts a first-order transition because of 
the presence of a cubic term in the LGW Hamiltonian.

Although we have no evidence of the existence of a CP$^{N-1}$ fixed
point for $N=3$, as claimed in Refs.~\cite{NCSOS-11,NCSOS-13}, our
results do not exclude it either.  Indeed, it is {\em a priori}
possible that the CP$^2$ lattice model we consider is outside the
attraction domain of the stable fixed point that controls the critical 
behavior of loop models~\cite{NCSOS-11,NCSOS-13}.  Clearly,
simulations of the loop model on larger lattices with a more careful
choice of the observables are needed to confirm that the loop model
has indeed a continuous transition and to exclude that the observed
behavior is a small-size apparent scaling behavior analogous to that
we have observed for $N\lesssim 50$.

We have also investigated the behavior of CP$^{N-1}$ models in the
large-$N$ limit.  In this case, depending on the specific Hamiltonian,
it is possible to observe both first-order and continuous
transitions. The latter ones can be obtained without any particular
tuning of the Hamiltonian parameters, therefore confirming the
existence of a large-$N$ stable fixed point.

Summarizing, we can conclude that the usual LGW picture
is not always able
to provide the correct qualitative description of the
finite-temperature transitions occurring in 3D CP$^{N-1}$ lattice
models. This conclusion is quite robust in the large-$N$ limit, as it
is obtained by analytic theoretical arguments.  In this regime the
gauge degrees of freedom must play a role, as it should be also
expected on the basis of the equivalence of the CP$^{N-1}$ field
theory with the abelian Higgs model that describes a dynamical U(1)
gauge field interacting with a U($N$) matter scalar field. For small
values of $N$, the existence of a 3D CP$^{N-1}$ universality class is
less robust, as it only relies on the numerical result of
Refs.~\cite{NCSOS-11,NCSOS-13} for $N=3$.  For the lattice model we
consider, there is no evidence of a critical transition. Clearly
additional numerical investigations are needed to settle the question.

In view of the equivalence of the abelian Higgs model with the
CP$^{N-1}$ model for large values of $N$ \cite{MZ-03}, it is also
worthwhile to discuss the critical behavior of the corresponding
lattice model. Its simplest version is given by the Hamiltonian
\begin{equation}
H = H_{\rm gauge} + H_\lambda,
\label{Abelian-Higgs}
\end{equation}
where $H_\lambda$ is the CP$^{N-1}$ Hamiltonian (\ref{hlg}) and 
\begin{equation}
H_{\rm gauge} = 
   c_{g} \sum_{{\bm x}\mu\nu} 
\left( \lambda_{{\bm x},\mu} \lambda_{{\bm x}+\hat{\mu},\nu} 
       \bar\lambda_{{\bm x}+\hat{\nu},\mu} \bar\lambda_{{\bm x},\nu} + 
       \hbox{c.c.} \right),
\end{equation}
is the usual Wilson Hamiltonian for the gauge field. This theory has
been extensively studied for $N=2$ \cite{TIM-05,TIM-06} and $N=4$
\cite{KHI-11}. In particular, for $N=4$, the numerical results of
Ref.~\cite{KHI-11} show that the first-order transition becomes weaker
by increasing the coupling $c_{g}$, and turns apparently into a
continuous one. A similar scenario may emerge for $N=3$.  It is
however, not obvious if these results are relevant for CP$^{N-1}$
models. Indeed, the equivalence of the abelian Higgs model
(\ref{Abelian-Higgs}) with the CP$^{N-1}$ has only been proved in the
large-$N$ limit. Therefore, we cannot take the results on the
existence of a continuous transition in the Abelian Higgs model as an
indication of the existence of a stable CP$^{N-1}$ fixed point.
Additional work is clearly needed.

\appendix
\section{Large-$N$ solution} \label{App-largeN}

In this Appendix we will solve the most general CP$^{N-1}$ theory in
the large-$N$ limit, generalizing to CP$^{N-1}$ the results of
Refs.~\cite{CP-02,SS-01}. We start from the general Hamiltonian
\begin{equation}
H = - N \sum_{\langle xy\rangle} 
   W(|\bar{\bm z}_y\cdot {\bm z}_x|^2),
\end{equation}
where $W(x)$ is an increasing function of $x$ for $0\le x \le 1$ to
guarantee ferromagnetism (therefore $W'(x) > 0$), and the sum is over
all links $\langle xy\rangle$ of a lattice. For definiteness, we will
consider a $d$ dimensional (hyper)-cubic lattice of size $L$ with
periodic boundary conditions.

To linearize the dependence of the Hamiltonian on the spin variables
${\bm z}_x$, we introduce a set of auxiliary fields \cite{CP-02}.  To
each lattice link we associate the real fields $\rho_{xy}$ and
$\lambda_{xy}$, and a complex field $\sigma_{xy}$, while to each
lattice site we associate a real field $\mu_x$ \cite{footnote-largeN}.
Using the identities (note that $\rho_{xy}$ is integrated in the
interval $[0,1]$)
\begin{eqnarray}
&& e^{N \beta W(|\bar{\bm z}_y\cdot {\bm z}_x|^2)} \propto \int d\rho_{xy}
d\lambda_{xy}\, 
\nonumber \\
&& \quad   \exp \left [N\beta \lambda_{xy} 
     (|\bar{\bm z}_y\cdot {\bm z}_x|^2 - \rho_{xy}) + N\beta W(\rho_{xy})\right],
\nonumber \\
&& e^{N\beta \lambda_{xy} |\bar{\bm z}_y\cdot {\bm z}_x|^2} \propto 
\int d\sigma_{xy} d\bar\sigma_{xy} \,
\nonumber \\
&& \quad   \exp\left
   [- N \beta \lambda_{xy}(|\sigma_{xy}|^2 - 
    \sigma_{xy} \bar{\bm z}_y\cdot {\bm z}_x - 
    \bar\sigma_{xy} {\bm z}_y\cdot \bar{\bm z}_x \right],
\nonumber \\ 
&& 
\delta (|z_x|^2 - 1) \propto  \int d\mu_x \, \exp[-N\beta (|z_x|^2 - 1) \mu_x],
\end{eqnarray}
we can rewrite the partition function as 
\begin{equation}
  Z = \int \prod_{\langle xy \rangle} 
   d\rho_{xy} d\lambda_{xy} d\sigma_{xy} d\bar\sigma_{xy} \,
    \prod_x d\mu_x d{\bm z}_x d\bar{\bm z}_x \, e^{N\beta A},
\end{equation}
where 
\begin{eqnarray}
A &=& -\sum_{\langle xy\rangle } \left[ 
   \lambda_{xy}(|\sigma_{xy}|^2 - \sigma_{xy} \bar{\bm z}_y\cdot
{\bm z}_x - \bar\sigma_{xy} {\bm z}_y\cdot
\bar{\bm z}_x ) \right.
\nonumber \\ 
&& \quad + \left. \lambda_{xy} \rho_{xy}  - W(\rho_{xy})\right] - 
  \sum_x (|z_x|^2 - 1) \mu_x \; .
\end{eqnarray}
We perform a saddle-point expansion writing 
\begin{eqnarray}
   \lambda_{xy} &=& \alpha + \hat{\lambda}_{xy}, \nonumber  \\
   \rho_{xy} &=& \tau + \hat{\rho}_{xy}, \nonumber  \\
   \mu_{xy} &=& \gamma + \hat{\mu}_{xy}, \nonumber  \\
   \sigma_{xy} &=& \delta + \hat{\sigma}_{xy} .
\end{eqnarray}
We first consider the case $\alpha \delta = 0$. In this case,
we have 
\begin{equation}
\langle \bar{\bm z}_x \cdot {\bm z}_y \rangle = {1\over \gamma \beta}
\delta_{xy}.
\label{random-noise}
\end{equation}
For $\alpha = 0$, the saddle-point equations give $W'(\tau) = 0$,
which is in contrast with our assumption that $W(x)$ is an increasing
function of $x$. We thus assume that $\alpha \not=0$ and $\delta = 0$.
The saddle-point equations give $\tau = 0$, $\alpha = W'(0)$, and
$\gamma \beta = 1$.  At the saddle point we have simply
\begin{equation}
   {\beta A \over L^d} = {\cal A}_1 = \beta d W(0) + 1.
\label{RN-A}
\end{equation}
We will call this case the white-noise solution, because of the 
absence of correlations, see Eq.~(\ref{random-noise}). 

If $\alpha,\delta \not= 0$, we can assume, using gauge invariance,
that $\delta$ is real. If we define $m_0^2 = (\gamma - 2 d \delta
\alpha)/\delta\alpha$ we obtain
\begin{equation}
\langle \bar{\bm z}_x \cdot {\bm z}_y \rangle = 
   {1\over \beta \delta \alpha} L^{-d} \sum_p
    {e^{ip(x-y)}\over \hat{p}^2 + m_0^2},
\end{equation}
where $\hat{p}_\mu = 2 \sin p_\mu/2$ and $\hat{p}^2 = \sum_\mu \hat{p}_\mu^2$.
We define 
\begin{eqnarray}
I(m_0^2,L) &=& {1\over L^{d}} \sum_p {1\over \hat{p}^2 + m_0^2}, \\
J(m_0^2,L) &=& {1\over L^{d}} \sum_p {\cos p_x\over \hat{p}^2 + m_0^2} 
\nonumber \\ 
   & = & {1\over 2d} [(2 d + m_0^2) I(m_0^2,L) - 1]. \nonumber 
\end{eqnarray}
The saddle-point equations become 
\begin{eqnarray}
 && 1 = {1\over \beta \delta \alpha} I(m_0^2,L), \nonumber \\
 && \alpha - W'(\tau) = 0, \nonumber \\
 && \tau + \delta^2 - {2\over \beta \alpha} J(m_0^2,L) = 0, \nonumber \\
 && \delta - {1\over \delta \beta \alpha} J(m_0^2,L) = 0,
\end{eqnarray}
which give 
\begin{eqnarray}
\delta &=& \sqrt{\tau}, \nonumber \\
\alpha &=& W'(\tau), \nonumber \\
\tau &=& \left({J(m_0^2,L) \over I(m_0^2,L)}\right)^2, \nonumber \\
\beta   &=& {I(m_0^2,L)\over \sqrt{\tau} W'(\tau)}.
\label{saddle2}
\end{eqnarray}
The quantity $\beta A$ is given by
\begin{equation}
{\beta A \over L^d} =
   - \beta d [\alpha \delta^2 + \alpha \tau - W(\tau)] + \beta\gamma -
    g(m_0^2,L) - \ln (\beta \delta \alpha),
\end{equation}
where
\begin{equation}
g(m_0^2,L) = L^{-d} \sum_p \ln (\hat{p}^2 + m_0^2).
\end{equation}
Using the gap equation, we can rewrite it as 
\begin{equation}
{\beta A \over L^d} = {\cal A}_2(m_0^2,L) = 
   \beta d W(\tau) + F(m_0^2,L),
\label{A2}
\end{equation}
where 
\begin{eqnarray}
F(m_0^2,L) &=& (m_0^2 + 2 d) I(m_0^2,L) - 2 d J(m_0^2,L) 
    \nonumber \\ 
  && - g(m_0^2,L) - \ln I(m_0^2,L).
\end{eqnarray}
To identify the correct phase, for each $\beta$, we should determine
the saddle point that maximizes $A$. Note that $F(m_0^2,L)$ is always
an increasing function of $m_0^{-2}$ that takes the value 1 for
$m_0=\infty$. Given that $W(\tau) > W(0)$, since $W(x)$ is an
increasing function of $x$, we obtain ${\cal A}_2(m_0^2,L) > {\cal
  A}_1$ for any finite value of $m_0^2$. Therefore, the solution of
Eq.~(\ref{saddle2}), if it exists, is always more stable than the
white-noise one.

We will now focus on the three-dimensional case. As discussed in
Ref.~\cite{SS-01}, one should take the infinite-volume limit of
$I(m_0^2,L)$ carefully. For $m_0^2 \not = 0$, the limit $L\to \infty$
is simply
\begin{equation}
I_\infty(m_0^2) = \int {d^3p\over (2\pi)^3} {1\over \hat{p}^2 + m_0^2},
\end{equation}
which varies between 0 (for $m_0^2 = +\infty$) and a maximum value 
\cite{GZ-77}
\begin{eqnarray}
&& I^* = I_\infty(0)  \\
&& \quad = {\sqrt{6}\over 192 \pi^3} 
\Gamma\left({1\over 24}\right) 
\Gamma\left({5\over 24}\right) 
\Gamma\left({7\over 24}\right) 
\Gamma\left({11\over 24}\right) .
\nonumber 
\end{eqnarray}
Numerically, $I^* \approx 0.252731.$ For $m_0^2 = 0$ we should take
into account the presence of a diverging zero mode. Therefore, for
$L\to\infty$ and $m_0^2\to 0$, we have
\begin{equation}
I(m_0^2,L) = L^{-d} \sum_{p\not=0} {1\over \hat{p}^2 + m_0^2} + {1\over L^3
m_0^2} \approx I^* + {1\over L^3 m_0^2}.
\end{equation}
Thus, for any $K > I^*$, there is always a solution $m_0(L)$ of the equation 
$I(m_0(L)^2,L) = K$, with $m_0(L)^2 \sim L^{-3}$. Thus, 
in three dimensions there
is a condensate phase that we can access by considering $m_0^2 \to 0$ and 
$L\to\infty$ at fixed $\Delta^{-1} =  m_0(L)^2 L^3$. In this limit 
$I(m_0^2,L)$ converges to $I^* + \Delta$. Note \cite{SS-01} 
that the zero-mode is irrelevant
for $g(m_0^2,L)$. If $m_0^2 \to 0$ and 
$L\to\infty$ at fixed $\Delta^{-1} =  m_0(L)^2 L^3$ we simply obtain 
\begin{equation}
g(m_0^2,L) \to g_\infty(0) = \int {d^3p\over (2\pi)^3} \ln \hat{p}^2.
\end{equation}
The presence of the zero mode implies that, for $L\to \infty$, all
quantities become functions of $m_0^2$ or $\Delta$, depending on how
the limit is taken.  With an abuse of notation, we will replace the
$L$ dependence with $\Delta$ in the following, to underscore that all
quantities are either functions of $m_0^2$ (and in this case $\Delta =
0$), or of $\Delta$ (in this case $m_0^2 = 0$).

We will now show that for $\beta$ small enough, there is no solution
of the saddle-point equations (\ref{saddle2}), provided that
$W'(\tau)$ is finite for $0 \le \tau \le 1$.  Indeed, as $m_0^2$ and
the condensate value $\Delta$ vary, $\beta$, computed from
Eq.~(\ref{saddle2}), is always larger than a specific value
$\beta_{\rm min}$. Therefore, for $\beta < \beta_{\rm min}$ the
white-noise solution is the relevant one.  To prove this point
consider the saddle-point equations (\ref{saddle2}).  The equation for
$\beta$ can be rewritten as
\begin{equation}
\beta = {H(m_0^2,\Delta)\over W'(\tau)} \qquad 
   H(m_0^2,\Delta) = {I(m_0^2,\Delta)^2 \over J(m_0^2,\Delta)}.
\label{saddle-point2}
\end{equation}
Expressing $J(m_0^2,\Delta)$ in terms of $I(m_0^2,\Delta)$, 
one can easily verify that:
\begin{itemize}
\item[i)] for $m_0^2 \to \infty$, the function $H(m_0^2,\Delta=0)$ converges to 1;
\item[ii)] as $m_0^2$ decreases, also $H(m_0^2,\Delta=0)$ decreases; the function
keeps decreasing also in the condensate phase up to $\Delta = \Delta_{\rm min}
= 1/3 - I^*$,
where is assumes the value $H_{\rm min} = H(0,\Delta_{\rm min}) =2/3$;
\item[iii)] for $\Delta > \Delta_{\rm min}$, $H(0,\Delta)$ increases, 
going to infinity as $\Delta \to \infty$.
\end{itemize} 
As for $\tau$, it increases from zero ($m_0^2\to\infty$) to 1
($m_0^2=0$, $\Delta \to \infty$). If $M$ is the maximum value that
$W'(\tau)$ takes in $0 \le \tau \le 1$, we immediately verify that
$\beta$, computed from (\ref{saddle-point2}) can never become smaller
than $2/(3 M)$.  In the interval $0 \le \beta < \beta_{\rm min}$ the
relevant solution is the white-noise one, coresponding to $m_0^2 =
\infty$.  Instead, for $\beta > \beta_{\rm min}$, the stable phase
corresponds to a solution of Eq.~(\ref{saddle2}), since the free
energy densities satisfy ${\cal A}_2(m_0^2,L) > {\cal
  A}_1$. Therefore, the point $\beta = \beta_{\rm min}$ is a
first-order transition point, where $m_0$ changes discontinuously. For
$\beta > \beta_{\rm min}$ there are may be several solutions of
Eq.~(\ref{saddle2}).  If, furthermore, we assume that also $W'(\tau)$
is an increasing function of $\tau$, we can easily conclude that the
stable solution for $\beta \ge \beta_{\rm min}$ always corresponds to
$m_0^2 = 0$ and $\Delta > 0$. At the first-order transition there is
coexistence between a disordered phase with $1/ m_0 = 0$ (the
correlation length vanishes) and an ordered phase with $m_0 = 0$ and
$\Delta > 0$.

It is interesting to note that the previous argument does not apply to
the gauge CP$^{N-1}$ Hamiltonian for which
\begin{equation}
H = - {1\over \beta} \sum_{\langle xy\rangle} 
    \ln I_0(2 N \beta |\bar{\bm z}_y\cdot {\bm z}_x|).
\end{equation}
In the large-$N$ limit it is equivalent to 
\begin{equation}
 H \approx - 2 N \sum_{\langle xy\rangle} |\bar{\bm z}_y\cdot {\bm z}_x|,
\end{equation}
so that $W(\tau) = 2 \sqrt{\tau}$ and $W'(\tau) = 1/\sqrt{\tau}$. The
derivative $W'(\tau)$ is clearly not bounded in $[0,1]$ and therefore
the previous argument does not apply.  For this model the gap equation
(\ref{saddle2}) becomes simply
\begin{equation}
\beta = I(m_0^2,L).
\end{equation}
For this Hamiltonian there is only a continuous transition for $m_0 = 0$. 

We can consider Hamiltonians that interpolate between Hamiltonians
(\ref{hcpn}) and (\ref{hlg}). For instance, we can consider
\begin{equation}
   W(\tau) = a \tau + 2 (1 - a) \sqrt{\tau}.
\end{equation}
In this case we observe a first-order transition for $0.7582\approx a
\le 1$.  In the opposite case there is a transition for $m_0=0$. The
first-order transition is however different from the one we discussed
before, as here the high-temperature coexisting phase corresponds to a
finite nonzero value of $m_0^2$.

To conclude, let us note that, in the derivation of the large-$N$
solution, one explicitly breaks gauge invariance, which is forbidden
by Elitzur's theorem \cite{Elitzur-75}. It is usually assumed this to
be only a technical problem.  Results for gauge-invariant quantities
are expected to be correct. This has been extensively checked in two
dimensions \cite{CR-93}.


\begin{thebibliography}{99}
\bibitem{footnoteMC} To update each lattice variable, we randomly
  choose either a standard Metropolis update, which ensures
  ergodicity, or a microcanonical move, which is more efficient than
  the Metropolis one but does not change the energy. Typically, on
  average we perform four/three microcanonical updates for every
  Metropolis proposal.  In the Metropolis update, changes are tuned so
  that the acceptance is approximately 1/3.
\bibitem{footnote-largeN} To be precise, the integrations over
  $\lambda_{xy}$ and $\mu_x$ are performed along a line in the complex
  plane parallel to the imaginary axis. What we mean here by real
  field, is that the integration is over a one-dimensional line that
  can be parametrized by a single real number. On the other hand, we
  must integrate $\sigma_{xy}$ over the whole complex plane, which
  explains why we name it a complex field.
\end{thebibliography}
\end{document}